\documentclass[fleqn,10pt]{wlscirep}
\usepackage[utf8]{inputenc}
\usepackage[T1]{fontenc}
\usepackage{algorithmic}
\usepackage[ruled,vlined]{algorithm2e}
\usepackage{acronym}
\usepackage{amsmath}
\usepackage{tabu}
\usepackage{multirow}

\title{Adaptive Extreme Edge Computing for Wearable Devices}

\author[1,*]{Erika Covi}
\author[2,*]{Elisa Donati}
\author[3,*]{Hadi Heidari}
\author[4,*]{David Kappel}
\author[3,*]{Xiangpeng Liang}
\author[2,*]{Melika Payvand}
\author[5,*]{Wei Wang}

\affil[1]{NaMLab gGmbH, N{\"o}thnitzer Strasse 64 a, 01187 Dresden, Germany}
\affil[2]{Institute of Neuroinformatics, University of Zurich and ETH Zurich, Switzerland}
\affil[3]{Microelectronics Lab (meLAB),  James Watt School of Engineering, University of Glasgow, G12 8QQ, UK}
\affil[4]{Bernstein Center for Computational Neuroscience, III Physikalisches Institut - Biophysik, Georg-August Universit{\"a}t, G{\"o}ttingen, Germany}

\affil[5]{The Andrew and Erna Viterbi Department of Electrical Engineering, Technion - Israel Institute of Technology, Haifa 32000, Israel,Formerly with Dipartimento di Elettronica, Informazione e Bioingegneria (DEIB), Politecnico di Milano and IU.NET, Milan, Italy}

\affil[*]{All authors contributed equally to this work}

\begin{abstract}

Wearable devices are a fast-growing technology with impact on personal healthcare for both society and economy. Due to the widespread of sensors in pervasive and distributed networks, power consumption, processing speed, and system adaptation are vital in future smart wearable devices. The visioning and forecasting of how to bring computation to the edge in smart sensors have already begun, with an aspiration to provide adaptive extreme edge computing. Here, we provide a holistic view of hardware and theoretical solutions towards smart wearable devices that can provide guidance to research in this pervasive computing era. We propose various solutions for biologically plausible models for continual learning in neuromorphic computing technologies for wearable sensors. To envision this concept, we provide a systematic outline in which prospective low power and low latency scenarios of wearable sensors in neuromorphic platforms are expected. We successively describe vital potential landscapes of neuromorphic processors exploiting complementary metal-oxide semiconductors (CMOS) and emerging memory technologies (e.g. memristive devices). Furthermore, we evaluate the requirements for edge computing within wearable devices in terms of footprint, power consumption, latency, and data size. We additionally investigate the challenges beyond neuromorphic computing hardware, algorithms and devices that could impede enhancement of adaptive edge computing in smart wearable devices.
\end{abstract}

\begin{document}
\acrodef{ADC}[ADC]{Analog to Digital Converter}
\acrodef{ADEXP}[AdExp-I\&F]{Adaptive-Exponential Integrate and Fire}
\acrodef{AER}[AER]{Address-Event Representation}
\acrodef{AEX}[AEX]{AER EXtension board}
\acrodef{AE}[AE]{Address-Event}
\acrodef{AFM}[AFM]{Atomic Force Microscope}
\acrodef{AGC}[AGC]{Automatic Gain Control}
\acrodef{AMDA}[AMDA]{AER Motherboard with D/A converters}
\acrodef{ANN}[ANN]{Artificial Neural Network}
\acrodef{API}[API]{Application Programming Interface}
\acrodef{ARM}[ARM]{Advanced RISC Machine}
\acrodef{ASIC}[ASIC]{Application Specific Integrated Circuit}
\acrodef{AdExp}[AdExp-IF]{Adaptive Exponential Integrate-and-Fire}
\acrodef{Backprop}[Backprop]{backpropagation}
\acrodef{BCM}[BMC]{Bienenstock-Cooper-Munro}
\acrodef{BD}[BD]{Bundled Data}
\acrodef{BEOL}[BEOL]{Back-end of Line}
\acrodef{BG}[BG]{Bias Generator}
\acrodef{BMI}[BMI]{Brain-Machince Interface}
\acrodef{BTB}[BTB]{band-to-band tunnelling}
\acrodef{BIS}[BIS]{Bioimpedance spectroscopy}
\acrodef{CAD}[CAD]{Computer Aided Design}
\acrodef{CAM}[CAM]{Content Addressable Memory}
\acrodef{CAVIAR}[CAVIAR]{Convolution AER Vision Architecture for Real-Time}
\acrodef{CA}[CA]{Cortical Automaton}
\acrodef{CCN}[CCN]{Cooperative and Competitive Network}
\acrodef{CDR}[CDR]{Clock-Data Recovery}
\acrodef{CFC}[CFC]{Current to Frequency Converter}
\acrodef{CHP}[CHP]{Communicating Hardware Processes}
\acrodef{CMIM}[CMIM]{Metal-insulator-metal Capacitor}
\acrodef{CML}[CML]{Current Mode Logic}
\acrodef{CMOL}[CMOL]{Hybrid CMOS nanoelectronic circuits}
\acrodef{CMOS}[CMOS]{Complementary Metal-Oxide-Semiconductor}
\acrodef{CNN}[CNN]{Convolutional Neural Network}
\acrodef{COTS}[COTS]{Commercial Off-The-Shelf}
\acrodef{CPG}[CPG]{Central Pattern Generator}
\acrodef{CPLD}[CPLD]{Complex Programmable Logic Device}
\acrodef{CPU}[CPU]{Central Processing Unit}
\acrodef{CSM}[CSM]{Cortical State Machine}
\acrodef{CSP}[CSP]{Constraint Satisfaction Problem}
\acrodef{CV}[CV]{Coefficient of Variation}
\acrodef{DAC}[DAC]{Digital to Analog Converter}
\acrodef{DAS}[DAS]{Dynamic Auditory Sensor}
\acrodef{DAVIS}[DAVIS]{Dynamic and Active Pixel Vision Sensor}
\acrodef{DBN}[DBN]{Deep Belief Network}
\acrodef{DFA}[DFA]{Deterministic Finite Automaton}
\acrodef{DIBL}[DIBL]{drain-induced-barrier-lowering}
\acrodef{DI}[DI]{delay insensitive}
\acrodef{DMA}[DMA]{Direct Memory Access}
\acrodef{DNF}[DNF]{Dynamic Neural Field}
\acrodef{DNN}[DNN]{Deep Neural Network}
\acrodef{DOF}[DOF]{Degrees of Freedom}
\acrodef{DPE}[DPE]{Dynamic Parameter Estimation}
\acrodef{DPI}[DPI]{Differential Pair Integrator}
\acrodef{DRRZ}[DR-RZ]{Dual-Rail Return-to-Zero}
\acrodef{DRAM}[DRAM]{Dynamic Random Access Memory}
\acrodef{DR}[DR]{Dual Rail}
\acrodef{DSP}[DSP]{Digital Signal Processor}
\acrodef{DVS}[DVS]{Dynamic Vision Sensor}
\acrodef{DYNAP}[DYNAP]{Dynamic Neuromorphic Asynchronous Processor}
\acrodef{EBL}[EBL]{Electron Beam Lithography}
\acrodef{EDVAC}[EDVAC]{Electronic Discrete Variable Automatic Computer}
\acrodef{ECG}[ECG]{Electrocardiography}
\acrodef{EMG}[EMG]{Electromyography}
\acrodef{EEG}[EEG]{Electroencephalography}
\acrodef{EIN}[EIN]{Excitatory-Inhibitory Network}
\acrodef{EM}[EM]{Expectation Maximization}
\acrodef{EOG}[EOG]{Electrooculography}
\acrodef{EPSC}[EPSC]{Excitatory Post-Synaptic Current}
\acrodef{EPSP}[EPSP]{Excitatory Post-Synaptic Potential}
\acrodef{EZ}[EZ]{Epileptogenic Zone}
\acrodef{FDSOI}[FDSOI]{Fully-Depleted Silicon on Insulator}
\acrodef{FET}[FET]{Field-Effect Transistor}
\acrodef{FFT}[FFT]{Fast Fourier Transform}
\acrodef{FI}[F-I]{Frequency-Current}
\acrodef{FPGA}[FPGA]{Field Programmable Gate Array}
\acrodef{FR}[FR]{Fast Ripple}
\acrodef{FSA}[FSA]{Finite State Automaton}
\acrodef{FSM}[FSM]{Finite State Machine}
\acrodef{GIDL}[GIDL]{gate-induced-drain-leakage}
\acrodef{GOPS}[GOPS]{Giga-Operations per Second}
\acrodef{GPU}[GPU]{Graphical Processing Unit}
\acrodef{GUI}[GUI]{Graphical User Interface}
\acrodef{HAL}[HAL]{Hardware Abstraction Layer}
\acrodef{HFO}[HFO]{High Frequency Oscillation}
\acrodef{HH}[H\&H]{Hodgkin \& Huxley}
\acrodef{HMM}[HMM]{Hidden Markov Model}
\acrodef{HRS}[HRS]{High-Resistive State}
\acrodef{HR}[HR]{Human Readable}
\acrodef{HSE}[HSE]{Handshaking Expansion}
\acrodef{HW}[HW]{Hardware}
\acrodef{ICT}[ICT]{Information and Communication Technology}
\acrodef{IC}[IC]{Integrated Circuit}
\acrodef{IEEG}[iEEG]{intracranial electroencephalography}
\acrodef{IF2DWTA}[IF2DWTA]{Integrate \& Fire 2--Dimensional WTA}
\acrodef{IFSLWTA}[IFSLWTA]{Integrate \& Fire Stop Learning WTA}
\acrodef{IF}[I\&F]{Integrate-and-Fire}
\acrodef{IMU}[IMU]{Inertial Measurement Unit}
\acrodef{INCF}[INCF]{International Neuroinformatics Coordinating Facility}
\acrodef{INI}[INI]{Institute of Neuroinformatics}
\acrodef{IO}[I/O]{Input/Output}
\acrodef{IPSC}[IPSC]{Inhibitory Post-Synaptic Current}
\acrodef{IPSP}[IPSP]{Inhibitory Post-Synaptic Potential}
\acrodef{IP}[IP]{Intellectual Property}
\acrodef{ISI}[ISI]{Inter-Spike Interval}
\acrodef{IoT}[IoT]{Internet of Things}
\acrodef{JFLAP}[JFLAP]{Java - Formal Languages and Automata Package}
\acrodef{KNN}[KNN]{K-Nearest Neighbour}
\acrodef{LEDR}[LEDR]{Level-Encoded Dual-Rail}
\acrodef{LFP}[LFP]{Local Field Potential}
\acrodef{LLC}[LLC]{Low Leakage Cell}
\acrodef{LNA}[LNA]{Low-Noise Amplifier}
\acrodef{LPF}[LPF]{Low Pass Filter}
\acrodef{LRS}[LRS]{Low-Resistive State}
\acrodef{LSM}[LSM]{Liquid State Machine}
\acrodef{LTD}[LTD]{Long Term Depression}
\acrodef{LTI}[LTI]{Linear Time-Invariant}
\acrodef{LTP}[LTP]{Long Term Potentiation}
\acrodef{LTU}[LTU]{Linear Threshold Unit}
\acrodef{LUT}[LUT]{Look-Up Table}
\acrodef{LVDS}[LVDS]{Low Voltage Differential Signaling}
\acrodef{MCMC}[MCMC]{Markov-Chain Monte Carlo}
\acrodef{MEMS}[MEMS]{Micro Electro Mechanical System}
\acrodef{MFR}[MFR]{Mean Firing Rate}
\acrodef{MIM}[MIM]{Metal Insulator Metal}
\acrodef{MLP}[MLP]{Multilayer Perceptron}
\acrodef{MOSCAP}[MOSCAP]{Metal Oxide Semiconductor Capacitor}
\acrodef{MOSFET}[MOSFET]{Metal Oxide Semiconductor Field-Effect Transistor}
\acrodef{MOS}[MOS]{Metal Oxide Semiconductor}
\acrodef{MRI}[MRI]{Magnetic Resonance Imaging}
\acrodef{NDFSM}[NDFSM]{Non-deterministic Finite State Machine} 
\acrodef{ND}[ND]{Noise-Driven}
\acrodef{NEF}[NEF]{Neural Engineering Framework}
\acrodef{NHML}[NHML]{Neuromorphic Hardware Mark-up Language}
\acrodef{NIL}[NIL]{Nano-Imprint Lithography}
\acrodef{NMDA}[NMDA]{N-Methyl-D-Aspartate}
\acrodef{NLP}[NLP]{Natural Language Processing}
\acrodef{NME}[NE]{Neuromorphic Engineering}
\acrodef{NN}[NN]{Neural Network}
\acrodef{NRZ}[NRZ]{Non-Return-to-Zero}
\acrodef{NSM}[NSM]{Neural State Machine}
\acrodef{OR}[OR]{Operating Room}
\acrodef{OTA}[OTA]{Operational Transconductance Amplifier}
\acrodef{PCB}[PCB]{Printed Circuit Board}
\acrodef{PCHB}[PCHB]{Pre-Charge Half-Buffer}
\acrodef{PCM}[PCM]{Phase Change Memory}
\acrodef{PE}[PE]{Phase Encoding}
\acrodef{PFA}[PFA]{Probabilistic Finite Automaton}
\acrodef{PFC}[PFC]{prefrontal cortex}
\acrodef{PFM}[PFM]{Pulse Frequency Modulation}
\acrodef{PR}[PR]{Production Rule}
\acrodef{PSC}[PSC]{Post-Synaptic Current}
\acrodef{PSP}[PSP]{Post-Synaptic Potential}
\acrodef{PSTH}[PSTH]{Peri-Stimulus Time Histogram}
\acrodef{PPG}[PPG]{Photoplethysmography}
\acrodef{QDI}[QDI]{Quasi Delay Insensitive}
\acrodef{RAM}[RAM]{Random Access Memory}
\acrodef{RDF}[RDF]{random dopant fluctuation}
\acrodef{RELU}[ReLu]{Rectified Linear Unit}
\acrodef{RLS}[RLS]{Recursive Least-Squares}
\acrodef{RMSE}[RMSE]{Root Mean Squared-Error}
\acrodef{RMS}[RMS]{Root Mean Squared}
\acrodef{RNN}[RNN]{Recurrent Neural Network}
\acrodef{ROLLS}[ROLLS]{Reconfigurable On-Line Learning Spiking}
\acrodef{RRAM}[R-RAM]{Resistive Random Access Memory}
\acrodef{R}[R]{Ripples}
\acrodef{SAC}[SAC]{Selective Attention Chip}
\acrodef{SAT}[SAT]{Boolean Satisfiability Problem}
\acrodef{SCX}[SCX]{Silicon CorteX}
\acrodef{SD}[SD]{Signal-Driven}
\acrodef{SEM}[SEM]{Spike-based Expectation Maximization}
\acrodef{SLAM}[SLAM]{Simultaneous Localization and Mapping}
\acrodef{SNN}[SNN]{Spiking Neural Network}
\acrodef{SNR}[SNR]{Signal to Noise Ratio}
\acrodef{SOC}[SOC]{System-On-Chip}
\acrodef{SOI}[SOI]{Silicon on Insulator}
\acrodef{SP}[SP]{Separation Property}
\acrodef{SRAM}[SRAM]{Static Random Access Memory}
\acrodef{STDP}[STDP]{Spike-Timing Dependent Plasticity}
\acrodef{STD}[STD]{Short-Term Depression}
\acrodef{STP}[STP]{Short-Term Plasticity}
\acrodef{STT-MRAM}[STT-MRAM]{Spin-Transfer Torque Magnetic Random Access Memory}
\acrodef{STT}[STT]{Spin-Transfer Torque}
\acrodef{SVM}[SVM]{Support Vector Machine}
\acrodef{SW}[SW]{Software}
\acrodef{TPU}[TPU]{Tensor Processing Unit}
\acrodef{TCAM}[TCAM]{Ternary Content-Addressable Memory}
\acrodef{TFT}[TFT]{Thin Film Transistor}
\acrodef{USB}[USB]{Universal Serial Bus}
\acrodef{VHDL}[VHDL]{VHSIC Hardware Description Language}
\acrodef{VLSI}[VLSI]{Very Large Scale Integration}
\acrodef{VOR}[VOR]{Vestibulo-Ocular Reflex}
\acrodef{WCST}[WCST]{Wisconsin Card Sorting Test}
\acrodef{WTA}[WTA]{Winner-Take-All}
\acrodef{XML}[XML]{eXtensible Mark-up Language}
\acrodef{CTXCTL}[CTXCTL]{CortexControl}
\acrodef{divmod3}[DIVMOD3]{divisibility of a number by three}
\acrodef{hWTA}[hWTA]{hard Winner-Take-All}
\acrodef{sWTA}[sWTA]{soft Winner-Take-All}
\acrodef{APMOM}[APMOM]{Alternate Polarity Metal On Metal}
\acrodef{VMM}[VMM]{Vector Matrix Multiplication}

\flushbottom
\maketitle

\section{Introduction}
\label{sec:intro}


Wearable devices can monitor various human body symptoms ranging from heart, respiration, movement, to brain activities. Such miniaturized devices using different sensors can detect, predict, and analyze the physical performance, physiological status, biochemical composition, and mental alertness of the human body. Despite advances in novel materials that can improve the resolution and sensitivity of sensors, modern wearable devices are facing various challenges such as low computing capability, high power consumption, high amount of data to be transmitted, and low speed of the data transmission. Conventional wearable sensing solutions mostly transmit the collected data to external servers for off-chip computing and processing. This approach typically creates an information bottleneck acting as one of the major limiting factors in lowering the power consumption and improving the speed of the operation of the sensing systems. In addition, the use of conventional remote servers with conventional signal processing techniques for processing these temporal real-time sensing data makes it computationally intensive and results in significant power consumption and hardware occupation. Moreover, standard von-Neumann architectures feature a physical separation between memory and processing unit, thus further increasing the power consumption to shuttle data between units. Such solutions always need a trade-off between power lifetime and computing capability. Bringing computing at the edge enables faster response times and opens the possibility of personalized always-on wearable devices able for continuously interacting and learning with the environment. However, a radical change of paradigm which uses innovative algorithms, circuits and memory devices is needed to maximize the system performance whilst keeping power and memory budgets at a minimum. 

Conventional computers, using Boolean and bit-precise digital representations and executing operations with time-multiplexed and clocked signal, are not optimized for fuzzy inputs and complex cognitive tasks such as pattern recognition, time series prediction, and decision making. Deep \acp{ANN} on the other hand have demonstrated amazing results in a wide range of pattern recognition tasks including machine vision, \ac{NLP}, and speech recognition~\cite{schmidhuber2015deep,LeCun2015}. Dedicated hardware \ac{ANN} accelerators, including \acp{GPU}, \acp{TPU}, and custom \acp{ASIC} with parallel architectures are being developed to execute these algorithms and obtain high accuracy inference results.
\acp{GPU} provide a substrate for parallel processing nature of the \acp{ANN} and thanks to its very long memory bus it is perfect for running \acp{VMM} which are at the core of the processing in deep neural networks.
Therefore, \acp{GPU} support the parallelism whose massive version exists in the brains for cognitive purposes, but they consume orders of magnitude more power than that of the brain~\cite{silver_etal16_alphago}, since they are clocked and the memory access is not localized. 
To solve this problem, \ac{ASIC} accelerators try to reduce the complexity of the structure by making the system more application specific and using clock gating and specific hardware structure which matches best to the structure of the mapped neural network to reduce power consumption through less memory read and data access~\cite{Chen2016,Cavigelli2016,Song2019,Lee2019}.

To go even further in power savings, there are two problems to be solved: (i) remove clock and (ii) perform computation with co-localization of memory and processor. The first problem calls for the development of event-based systems, where processing is performed ``asynchronously", i.e. only when there are input ``events". The algorithmic basis for this kind of ``asynchronous" processing is \ac{SNN}, in which neurons spike asynchronously only to communicate information to each other. 

To avoid the data movement between the memory and the processor, the memory element should be not only used to store data but also to perform computation inside the processor. This approach is called ``in-memory computing". These two approaches of (i) event-based systems and (ii) in-memory computing, together with (iii) massive parallelism, are the three fundamental principles which have led to the development of neuromorphic computing, and to the realization of highly efficient neuromorphic platforms~\cite{Furber2014,Merolla2014,Davies_etal18,Schemmel2010,Moradi_etal17,frenkel20180}. Therefore, in this article, we will refer to event-based highly parallel systems that are able to perform real-time sensory processing.

Despite that current fully \ac{CMOS} implementations of neuromorphic platforms have shown remarkable performance in terms of power efficiency and classification accuracy, there are still some bottlenecks hindering the design of embedded sensing and processing systems. First, the memory used is typically \ac{SRAM}, which has very low static power consumption, but it is a large element (6 transistors per cell) and it is volatile. The latter feature implies that the information about the network configuration has to be stored elsewhere and transferred to the system at its startup. For large networks, it may take tens of minutes before the system is ready for normal operation. Second, always-on adaptive systems need to work with time constants that have the same time-span of the task that is being learned (e.g. longer than seconds). Implementing such long time constants in neuromorphic \ac{CMOS} circuits is impractical, since it requires large area capacitors.

To overcome the limitations of fully \ac{CMOS}-based approaches, the intrinsic unique physical properties of emerging memristive devices can be exploited for both long-term (non-volatile) weight storage and short-term (volatile) task-relevant timescales. In particular, non-volatile devices feature retention times on a long time scale ($>$10 years,~\cite{cheng2012IEDM,udayakumar2013IMW,goux2014VLSI,golonzka2018IEDM}) while showing weight reconfigurability with voltages compatible with typical CMOS circuits ($\leq$ 3.3 V). Volatile devices, instead, can have time constants in the order of tens of milliseconds to seconds  (\cite{jo2015TED,yang2017IEDM,ZhongruiWang2017,WeiWang2019nc,wang2019IEDM,covi2019ICECS}), thus being able to emulate biological time constants. This non-volatile / volatile property of memristive devices, together with a small footprint and power efficiency, has indeed attracted a lot of interest in the last ten years~\cite{linaresbarranco2009NatPrec,Ielmini2018,chicca2020APL}. However, memristive technology has to be supported by \emph{ad hoc} theoretically sound biologically plausible algorithms enabling continual learning and capable to exploit the intrinsic physical properties of memristive devices, such as stochasticity, to achieve accuracy performance comparable to state-of-the-art \ac{ANN} whilst reducing the power consumption.

\begin{figure}
    \centering
    \includegraphics[width=\textwidth]{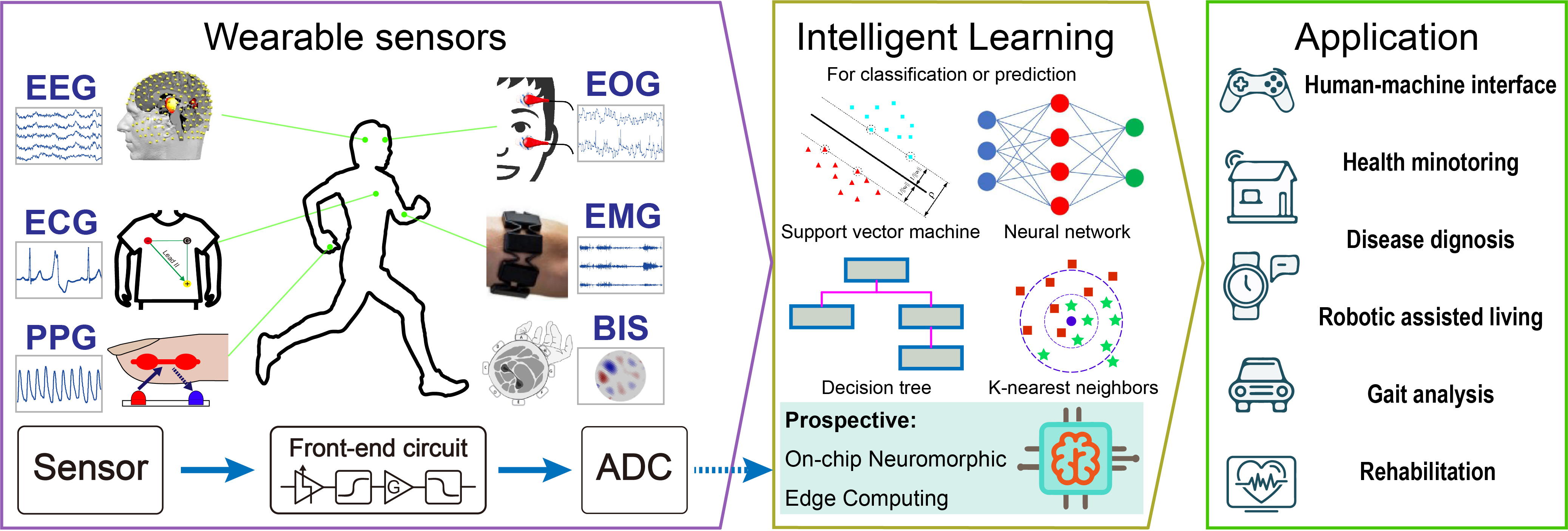}
    \caption{A graphical overview of adaptive edge computing in wearable biomedical devices. The figure shows the pathway from wearable sensors to their application through intelligent learning.}
    \label{fig:wearable}
\end{figure}

This review discusses the challenges to undertake for designing extreme edge computing wearable devices in four different categories: (i) the state-of-the-art wearable sensors and main restrictions towards low-power and high performance learning capabilities; (ii) different algorithms for modeling biologically plausible continual learning; (iii) CMOS-based neuromorphic processors and signal processing techniques enabling low-power local edge computing strategies; (iv) emerging memristive devices for more efficient and scalable embedded intelligent systems. As graphically summarized in Fig.~\ref{fig:wearable}, we argue that a holistic approach which combines and exploits all the strengths of these four categories in a co-designed system is the key factor enabling future generations of smart sensing systems.

\section{Wearable sensors}
\label{sec:wearable}

Sensors act as the information collector of a machine or a system that can respond to its physical ambient environment. They are able to translate a specific type of information from a physical environment such as the human body to an electrical signal (\cite{GaoWei2016}). For collecting the information from the human body environment, wearable versions of the machine or the system, i.e. wearable devices, would be of great convenient and helpful. Wearable devices require miniaturize, flexible, and highly sensitive sensors to capture clear information from the body. However, from processing aspect and to make a signal meaningful towards personalized devices, further development is still needed. 

Due to the fact that the sensing signal is relatively weak and noisy, a readout circuit (normally composed by an amplifier, a conditioning circuit and an analogue signal processing unit) is necessary to make the signal readable for a system (\cite{GaoWei2016, Kanoun2004}). The subsequent high-level system will process the data and send commands to actuators for a closed-loop control or interaction (\cite{Lopez2018, Nweke2018, Witkowski2014}). For various applications ranging from the human-machine interface (\cite{Lopez2018}) to health monitoring (\cite{RN62, RN18}), different combinations of sensor and system have been developed over the past decade (\cite{RN8, RN144}). The use of machine learning empowers sensor to build a novel smart application. The examples will be provided in the next section. 

\vspace{6mm}

\subsection{Wearable sensors with machine learning} 
Recently, the field of artificial intelligence further boosts the possibility of smart wearable sensory systems. The emerging intelligent applications and high-performance systems require more complexity and demand sensory units accurately describe the physical object. The decision-making unit or algorithm can therefore output a more reliable result (\cite{RN36, RN44, RN39, RN144, RN49}). Depending on the signal acquiring position, Fig.~\ref{fig:wearable} summaries the four biopotential sensors and two widely used wearable sensors along with their learning systems and applications. The sensors for the biopotential will be introduced first, and the other two wearable sensors will be provided separately.

The biopotential signal can be extracted from the human body using a sensor with direct electrode contact. The electrochemical activity of the cells in nervous, muscular and glandular tissue generates ionic currents in the body. An electrode-electrolyte transducer is needed to convert the ionic current to electric current for the front-end circuit. The electrode that is normally made up of mental can be oxidized by the electrolyte, generating metal ions and free electrons. In addition, the anions in the electrolyte can also be oxidized to neutral atoms and free electrons. These free electrons result in current flow through the electrode. Thus, the surface potential generated by the electrochemical activities in cells can be sensed by the electrode. However, the bio-signals sensed by the electrode are weak and noisy. Before digitizing the collected signals by analog-to-digital converter, an analogue front-end is essential to provide a readable signal. The design requirements of the front-end for the biopotential electrodes can be summarized as follow: i) high common mode rejection ratio; ii) high signal-to-noise-ratio; iii) low-power consumption; iv) signal filtering, and v) configurable gain (\cite{RN88}).


\textbf{\emph{\ac{ECG}.}} \ac{ECG} is the electrical activity generated by the electrochemistry around cardiac tissue. Containing morphological or statistical features, \ac{ECG} provides comprehensive information for analyzing and diagnosing cardiovascular diseases (\cite{RN72}). In the previous study, automatic \ac{ECG} classification has been achieved using machine learning algorithms, such as \ac{DNN} (\cite{RN74, RN45}), \ac{SVM} (\cite{RN34, RN84}), and \ac{RNN} (\cite{RN132, RN23}). According to Association for the Advancement of Medical Instrumentation, there are five classes of \ac{ECG} type of interest: normal, ventricular, supraventricular, fusion of normal and ventricular, and unknown beats. These methodologies can be evaluated by available \ac{ECG} database and yield over 90$\%$ accuracy and sensitivity for the five classes, which is essential for future cardiovascular health monitoring. In wearable application, \cite{RN75} and \cite{RN76} present systems that measure \ac{ECG} and send it to the cloud for classification and health monitoring. 


\textbf{\emph{\ac{EEG}.}} Our brain neurons communicate with each other through electrical impulses. An \ac{EEG} electrode can help to detect potential information associated with this activity through investigating \ac{EEG} (\cite{RN79, RN81}) in the surface of the skull. In comparison with other biopotential signals, surface \ac{EEG} is relatively weak (normally in the range of microvolt-level) and noisy (\cite{RN87, RN232}). Therefore, it requires high input impedance readout circuit and intensive signal pre-processing for clean \ac{EEG} data (\cite{RN79, RN88}). While wet-electrode (Ag/AgCl) 
is more precise and more suitable for clinical purpose, passive dry-electrode is more suitable for daily health monitoring and brain-machine interface (\cite{RN87, RN80}). Besides, the applications also include mental disorder (\cite{RN90}), driving safety (\cite{RN80, RN81}), and emotion evaluation (\cite{RN244}). 
A commercial biopotential data acquisition system, Biosemi Active Two, provides up to 256 channels for \ac{EEG} analysis (\cite{RN93}). For a specific application, we can reduce the number of electrodes to only detect the relevant areas, such as 19 channels for depression diagnosis (\cite{RN33}), four channels for evaluating driver vigilance (\cite{RN81}) and 64 channels for emotional state classification (\cite{RN244}). 
Although \ac{EEG} is on-body biopotential, most of the existing \ac{EEG} researches employed offline learning and analysis because of the system complexity and the high number of channels. In wearable real-time applications, usually a smaller number of channels were selected and the data were wirelessly sent to cloud for further processing (\cite{RN97, RN80, RN81, RN96}). 

\textbf{\emph{\ac{EOG}.}} The eye movement, which results in potential variations around eyes as \ac{EOG}, is a combined effect of environmental and psychological changes. It returns relatively weak voltage (0.01-0.1mV) and low frequency (0-10Hz) (\cite{RN232}). Differ from other eye tracking techniques using a video camera and infrared, \ac{EOG} provides a lightweight, inexpensive and fully wearable solution to access human’s eye movement (\cite{RN238}). It is the most widely used approach of wearable human-machine interface, especially for assisting quadriplegics (\cite{RN238}). It has been used to control a wheelchair (\cite{RN239}), control a prosthesis limb (\cite{RN240}),(\cite{Witkowski2014}) evaluate sleeping (\cite{RN237, RN236, RN234}). Additionally, recent studies fuse \ac{EEG} and \ac{EOG} to increase the degree of freedom of signal and enhance the system reliability because their similar implicit information such as sleepiness (\cite{RN237, RN241}) and mental health (\cite{RN243}). \ac{EOG} can also act as a supplement to provide additional functions or commands to an \ac{EEG} system (\cite{RN233, RN242, Witkowski2014}).

\textbf{\emph{\ac{EMG}.}} \ac{EMG} is an electrodiagnostic method for recording and analyzing the electrical activity generated by skeletal muscles. 
\ac{EMG} is generated by skeletal muscle movement, which frequently occurs in arms and legs. It yields higher amplitude (up to 10 millivolts) and bandwidth (20-1000Hz) compared to the other biopotentials (\cite{RN232, RN88}). Near the active muscle, different oscillation signals can be measured by a dry electrode array, which allows the computer to sense and decode body motion (\cite{RN224, RN228, RN221}). A prime example is the Myo armband of Thalmic Labs, which is a commercial multi-sensor device that consists of \ac{EMG} sensors, gyroscope, accelerometer and magnetometer (\cite{RN222}). The sensory data is sent to phone or PC via Bluetooth, at which various body movements can be obtained by feature extraction and machine learning. Moreover, the application of \ac{EMG} is frequently linked to target control like a wheelchair (\cite{RN218}) and prosthetic hand (\cite{RN225, RN226}) for assisting disabled people. In addition, its application also includes sign language recognition (\cite{RN224}), diagnosis of neuromuscular disorders (\cite{RN228, RN227}), analysis of walking strides (\cite{RN221}) and virtual reality (\cite{RN229}). Machine learning enables the system to overcome the variation of \ac{EMG} signals from different users (\cite{RN224, RN228}). 

\textbf{\emph{\ac{PPG}.}} \ac{PPG} is an non-invasive and low-cost optical measurement method that is often used for blood pressure and heart rate monitoring in wearable devices. The optical properties in skin and tissue are periodically changes due to the blood flow driven by the heartbeat. By using a light emitter toward the skin surface, the photosensor can detect the variations in light absorption normally from wrist or finger. This variation signal is called \ac{PPG} which is highly relevant to the rhythm of the cardiovascular system (\cite{RN255}). Compared with \ac{ECG}, \ac{PPG} is easily accessible and low cost, which makes it an ideal intermedia of wearable heart rate measurement. The main disadvantage against \ac{ECG} is that the \ac{PPG} is not unique for different persons and body positions. Thus, further analysis of \ac{PPG} requires machine learning or other statistics tools for calibrating the signal to different scenarios. For example, it can be used in biometric identification after deep learning (\cite{RN252, RN256}). It is worth mentioning that \ac{PPG} is a strong supplementary in the application of \ac{ECG}. 

\textbf{\emph{\ac{BIS}.}} \ac{BIS} is another low-cost and powerful sensing technique that provides informative body parameters. The principle is that cell membrane behaves like a frequency-dependent capacitor and impedance. The emitter electrodes generate multifrequency excitation signal (0.1-100MHz) on the skin while the receiver electrodes collect these current for demodulating the impedance spectral data of the tissue in between (\cite{CAYTAK2019265, RN258}). Compared to homogeneous materials, body tissue presents more complicated impedance spectra because of the cell membranes and macromolecules. Therefore, the tissue conditions, such as muscle concentration, structural and chemical composition, can be analysed through BIS. The BIS can measure body composition such as fat and water (\cite{RN258}). Based on the different setup in terms of position and frequency, it can also be helpful in the early detection of diseases such as lymphedema, organ ischemia and cancer (\cite{RN259}). Furthermore, multiple pair-wise electrodes can form electrical impedance tomography that describes impedance distribution. By embedding these electrodes in a wristband, the tomography can estimate hand gesture after training, which is another novel solution of inexpensive human-machine interface (\cite{RN55}).

\subsection{Multisensory fusion in wearable devices}
\label{sec:sensor-fusion}

Every sensor has its own limitation. In some demanding cases, an individual sensor itself cannot satisfy the system requirement such as accuracy or robustness (\cite{RN2, RN1, RN10, RN144}). The solution involves increasing the number and type of sensors to form a multisensory system or sensor network for one measurement purpose(\cite{RN2, RN1, RN10}). Multiple types of sensor synergistically working in a system provide more dimensions of input to fully map an object onto the data stream. Different sensors return different data with respect to sampling rate, number of input and the information behind the data. Machine learning models, such as \ac{ANN} and \ac{SVM}, can be designed to combine multiple sources of data. Depended on the application, sensor types and data structure, several approaches have been proposed for multisensory fusion. Generally, in such a system, machine learning is frequently used and plays an vital role in merging different sources of sensory data based on its multidimensional data processing mechanism. The machine learning algorithms allow sensory fusion occurs at the signal, feature or decision level(\cite{RN1, RN10}). The results showed that a multisensory system is advantageous in improving system performance. For example, the fusion of \ac{ECG} and \ac{PPG} pattern can be an informative physiological parameter for robust medical assessment (\cite{RN253}). Counting the peak intervals between \ac{PPG} and \ac{ECG} can estimate the arterial blood pressure (\cite{RN254}). Interestingly, a recent study shows that the QRS complex of \ac{ECG} can be reconstructed from \ac{PPG} by a novel transformed attentional neural networks after training (\cite{RN257}). This could be beneficial for the accessibility of wearable \ac{ECG}.

\vspace{6mm}

\subsection{Challenges towards smart wearable sensors with edge computing}
Given the potential of the sensory system with machine learning, the main challenge raised is the shortage of power and computing efficient (\cite{Kanoun2004}). The novel applications using multiple sensors and high learning ability usually require more energy in the wearable computing unit (\cite{RN18}). Nevertheless, the power supply in the wearable domain is a difficulty with existing battery technologies. This weakness limits the further development of smart wearable device (\cite{RN18}). The existing solution is to wirelessly transfer the raw data onto a cloud where the computationally intensive algorithm is implemented (\cite{RN66}). However, this solution is not ideal considering 1) the complexity of using a wireless module, 2) the non-negligible power consumption, 3) the amount of data, 4) the space limitation due to the range of wireless transmission, 5) privacy issues due to the broadcast of signals, 6) non-negligible time latency due to communication channel. These drawbacks strongly limit the application of wearable sensors.

Implementation of \ac{ANN} in von Neumann architectures, which has been frequently used in sensors, is power-hungry. Conversely, it has been reported that signal processing activity in the brain is several orders of magnitudes more power-efficient and one order in processing rate better than digital systems (\cite{RN98}). 
Compared to conventional approaches based on a binary digital system, brain-inspired neuromorphic hardware yet to be advanced in the contexts of data storage and removal as well as their transmission between different units. In this perspective, a neuromorphic chip with a built-in intelligent algorithm can act as a front-end processor next to the sensor. The conventional \acp{ADC} could be replaced by a delta encoder or feature extractor converting the sensor analog output to spike-based signal for the hardware (see Section~\ref{sec:cmos}). In the end, the output becomes the result of recognition or prediction instead of an intensive data stream. 
In this way, the computation occurs at the local edge under low power and brain-like architecture.


\section{Models for biologically plausible continual learning}
\label{sec:models}
\begin{figure}
    \centering
    \includegraphics[width=\textwidth]{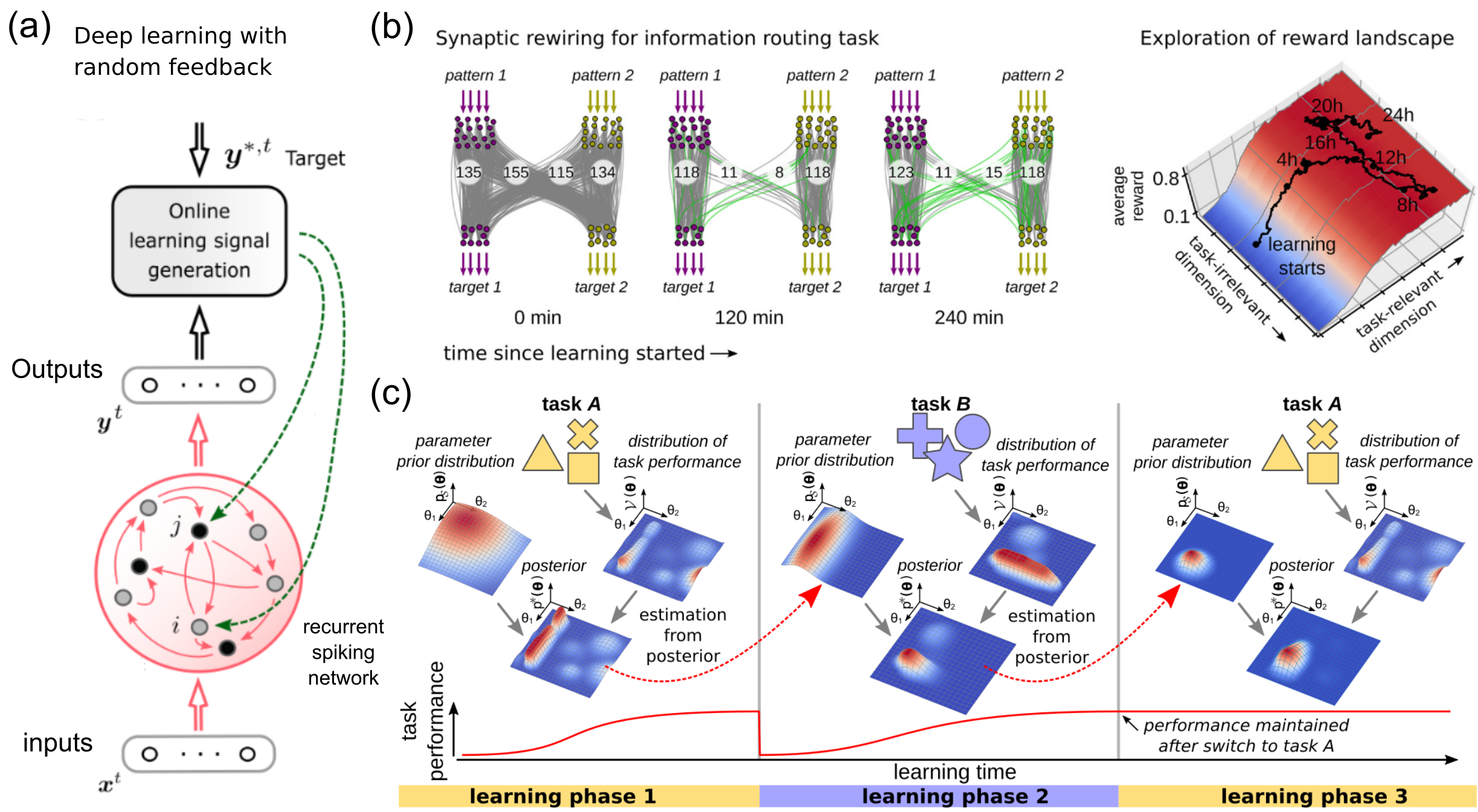}
    \caption{\textbf{Biologically inspired models of learning in spiking neural networks} (a) The e-prop algorithm \cite{bellec2019_eprop} approximates back-propagation through time using random feedback to propagate error signals to synapses of a recurrent SNN (adapted from \cite{bellec2020solution}) (b) Synaptic sampling \cite{kappel2015network} exploits the variability of learning rules and redundancy in the task solution space to learn sparse and robust network configurations (adapted from \cite{kappel2018dynamic}) (c) Overcoming forgetting by selectively slowing down weight changes \cite{KirkpatrickETAL:17}. After learning a first task A, parameter distributions are absorbed into a prior distribution that confines the motility of synaptic weights in subsequent tasks (task B).} 
    \label{fig:models}
\end{figure}

In this section we will highlight some recently introduced methods to port the power of modern machine learning to neuromorphic edge devices. In the last couple of years, machine learning has made big steps forward reaching close-to human performance on a wide range of tasks. Many of the most successful machine learning methods are based on artificial neural networks (ANN), which are inspired by the organization of information processing in the brain. However -- somewhat contradictory -- mapping modern ANN learning methods to brain-inspired hardware poses considerable challenges to the algorithm and hardware design. The main reason for this is, that the development of machine learning algorithms has been strongly influenced by the development of powerful mainframe computers that perform learning offline in big server farms only eventually sending back results to the user. While this development has paved the ground for today's success of ANNs, it has also lead the field away from following the principles used in biology for efficient learning. In the following Section~\ref{sec:bio-algorithms} we will review recent approaches to combine the strengths of modern machine learning and brain-inspired algorithms, that are of particular interest for edge computing applications. In Section~\ref{sec:pruning} we will focus on the problem to cope with extreme memory constraints by exploiting sparsity. In Section~\ref{sec:model-challenge} we will highlight additional open challenges and future work.
\vspace{8mm}

\subsection{Brain-inspired learning algorithms for neuromorphic hardware}
\label{sec:bio-algorithms}
Today, the dominating method for training artificial neural networks is the error \ac{Backprop} algorithm \cite{rumelhart1986learning}, which provides an efficient and scalable solution to adapting the network parameters to a set of training data. \ac{Backprop} is an iterative, gradient-based, supervised learning algorithm that operates in three phases. First, a given input activation is propagated through the network to generate the output based on the current set of parameters. Then, the mismatch between the generated outputs and target values is computed using a loss function, and propagated backwards through the network architecture to compute suitable weight changes. Finally, the network parameters are updated to reduce the loss. We will not go into the details behind \ac{Backprop} here, but see \cite{schmidhuber2015deep} for an excellent review and historical survey of the development of the algorithm. The problem of porting \ac{Backprop} to neuromorphic hardware stems form a well-known shortcoming of the algorithm known as \emph{locking} 
-- the weights of a network can only be updated after a full forwards propagation of the data through the network, followed by loss evaluation, then finally after waiting for the back-propagation of error gradients \cite{czarnecki2017understanding}. Locking prevents an efficient implementation of \ac{Backprop} on online distributed architectures. Also, \ac{Backprop} is not well suited for spiking neural networks which have non-differentiable output functions. These problems have been recently addressed in brain-inspired variants of the \ac{Backprop} algorithm.

\vspace{4mm}

\subsubsection{Brain-inspired alternatives to error backpropagation}

In recent years a number of methods have been proposed to approximate the gradient computation performed by \ac{Backprop} in order to prevent locking (see \cite{richards2019deep} for a recent review). \cite{lillicrap2016random,samadi2017deep} proposed to replace the non-local error back-propagating term of the \ac{Backprop} algorithm by sending the loss through a fixed feedback network with random weights that are excluded from training. In this approach, named \emph{random feedback alignment} the back-propagating error signal acts as a local feedback to each synapse, similar to a reward signal in reinforcement learning. The fixed random feedback network de-correlates the error signals providing individual feedback to each synapse. Lillicrap et al. could show that this simple approach already provides a viable approximation to the exact \ac{Backprop} algorithm and performs well for practical machine learning problems of moderate size. In \cite{neftci2017event} an event-based version of random feedback alignment, that is well suitable for neuromorphic hardware, was introduced. This approach was further generalized in \cite{payvand2019error} to include a larger class of algorithms that use error feedback signals.

An efficient model for learning complex sequences in spiking neural networks, named \emph{Superspike}, was introduced in  \cite{zenke2018superspike}. The model also uses a learning rule that is modulated by error feedback signals and locally minimizes the mismatch between the network output and a target spike train. To overcome the problem of non-differentiable output, Superspike uses a surrogate gradient approach that replaces the infinitely steep spike events with a finite auxiliary function at the time points of network spike events \cite{hinton2012neural,bengio2013estimating}. As in random feedback alignment, learning signals are communicated to the synapses via a feedback network with fixed weights. Using this approach Zenke and others could demonstrate efficient learning of complex sequences in spiking networks.
\vspace{4mm}

Another approach to approximate \ac{Backprop} in spiking neural networks uses an anatomical detail of Cortical neurons. \cite{sacramento2017dendritic} introduced a biologically inspired two-compartment neuron model that approximates the error backpropagation algorithm by minimizing a local dendritic prediction error. 
\cite{goltz2019fast} port learning by \ac{Backprop} to neuromorphic hardware by incorporating dynamics  with  finite  time  constants  and  by  optimizing  the backward  pass  with  respect  to  substrate  variability. They demonstrate the algorithm on the BrainScaleS analog neuromorphic architecture.

\vspace{8mm}

\subsubsection{Brain-inspired alternatives to backpropagation through time}
Recurrent neural network (RNN) architectures often show superior learning results for tasks that involve a temporal dimension, which is often the case for edge computing applications. Porting learning algorithms for RNNs is therefore of utmost importance for efficient machine learning on the edge. Backpropagation through time (BPTT) -- the standard RNN learning method used in most GPU implementations -- unfolds the network in time and keep this extended structure in memory to propagate information forward and backward which poses a severe challenge to the power and area constraints of edge computing. Recent theoretical results \cite{bellec2018long,bellec2019_eprop} show that the power of BPTT can be brought to biologically inspired spiking neural networks (SNN) while at the same time the unfolding can be prevented in an approximation that operates only forward in time, enabling \emph{online, always-on} learning. This algorithm operates at every synapse in parallel and incrementally updates the synaptic weights. As for random feedback alignment and Superspike discussed above, the weight update depends only on three factors, where the first two are determined by the states of the two related input/output neurons, and the third is given by synapse-specific feedback conveying the mismatch between the target and the actual output (see Fig.~\ref{fig:models}a for an illustration). The temporal gap between these factors is mitigated by an \emph{eligibility trace} describing a transient dynamic. Eligibility traces, have been theoretically predicted for a long time \cite{williams1992simple,izhikevich2007solving}, and have also recently been observed experimentally in the brain  \cite{yagishita2014critical,he2015distinct,brzosko2015retroactive,bittner2017behavioral}.

\vspace{8mm}

\subsection{Efficient learning under stringent memory constraints}
\label{sec:pruning}
The amount of available resources in neuromorphic systems is kept low to increase energy efficiency. Memory elements are especially impactful on the energy budget. Therefore, algorithms are needed that make efficient use of the available memory resources. The largest amount of memory in a network is usually consumed by the synaptic weights. Since in practice, the weights of many connections in a network converge to values close to zero, several methods have been proposed to reduce the memory footprint of machine learning algorithms by exploiting sparsity in the network connectivity. We will discuss here two types of algorithms: (1) those that are based on \emph{pruning connections after learning} and (2) \emph{online} learning with \emph{sparse} networks. These two types of sparse learning algorithms are discussed in the following sections.

\subsubsection{Pruning}
Many approaches to exploit sparsity in learning algorithms focus on pruning the network after training (see \cite{gale2019state} for a recent review). Simple methods rely on pruning by magnitude, simply by eliminating the weakest (closest to zero) weights in the network \cite{strom1997sparse,collins2014memory,han2015learning}. Some methods based on this idea have reported impressive sparsity rates of over 95\% for standard machine learning benchmarks with negligible performance loss \cite{guo2016dynamic,zhu2017prune}.
Other methods are based on theoretical motivations and classical sparsification and regularization techniques \cite{molchanov2017variational,louizos2017learning,ullrich2017soft}. These models reach high compression rates. \cite{dai2019nest} proposed a method to iteratively grow and prune a network in order to generate a compact yet precise solution. They provide a detailed comparison with state of the art dense networks and other pruning methods and reaching sparsity above 99\% for the LeNet-5 benchmark.\\

\subsubsection{Online learning in sparse networks}
A number of authors also introduced methods that work directly with sparse networks during training, which is often the more interesting case for neuromorphic applications with online training. \cite{bellec2017deep} introduced an algorithm for online stochastic rewiring in deep neural networks that works with a fixed number of synaptic connections throughout learning. The algorithm showed close-to state of the art performance at up to 98\% sparsity.
Sparse evolutionary training (SET) \cite{mocanu2018scalable} introduced a heuristic approach that prunes the smallest weights and regrows new weights in random locations. Dynamic Sparse Reparameterization \cite{mostafa2019parameter} introduces a prune-redistribute-regrowth cycle. They demonstrated compelling performance levels also for very deep neural network architectures. \cite{lee2018snip} introduced a single shot pruning algorithm that yields sparse networks based on a saliency criterion prior to the actual training. \cite{dettmers2019sparse} introduced a refined method for online pruning and redistribution that surpasses the previous methods in terms of sparsity and learning performance.

\vspace{4mm}
\subsection{Open challenges and future work}
\label{sec:model-challenge} 
As outlined above, edge computing poses quite specific challenges to learning algorithms that are substantially different from requirements of classical applications. Some of the algorithms outlined above have already been succesfully ported to neuromorphic hardware. For example, the e-prop algorithm of \cite{bellec2018long} has been implemented on the SpiNNaker~2 chip yielding an additional energy reduction by two orders of magnitude compared to a X86 implementation \cite{liu2018memory}. See the next Section~\ref{sec:cmos}
for more details on available neuromorphic hardware and their applications.

In the remainder of this section we will highlight open challenges that remain to be solved for efficient learning in edge computing applications. In addition to the stringent memory and power constraints learning at the edge also has to function in an online scenario where data arrive in a continuous stream. Some dedicated hardware resources, e.g. like memristive devices discussed in Section~\ref{sec:memristive}, may also show high levels in intrinsic variability, so the learning algorithm should be robust against these noise sources. In this section we discuss recent advances in this line of research and provide food for thought on how these specific challenges can be approached in future work.

\vspace{4mm}
\subsubsection{Fault-tolerant robust learning algorithms for neuromorphic devices}
Here we review recent advances in using inspiration from biology to make learning algorithms robust against device variability. Several authors have suggested that device noise and variability should not be seen as a nuisance, but rather can serve as a computational resource for network simulation and learning algorithms (see \cite{maass2014noise} for a thorough discussion). \cite{pecevski2016learning} have shown that variability in neuronal outputs can be exploited to learn complex statistical dependencies between sensory stimuli. The stochastic behavior of the neurons is used in this model to compute probabilistic inference, while biologically motivated learning rules, that only require local information at the synapses can be used to update the synaptic weights. A theoretical foundation of the model shows that the spiking network performs a Markov chain Monte Carlo sampling process, that allows the network to 'reason' about statistical problems.

This idea is taken one step further in \cite{neftci2015unsupervised} by showing that also the variability of synaptic transmission can be used for stochastic computing. The intrinsic noise of synaptic release is used to drive a sampling process. It was shown that this model can be implemented in an event-based fashion and was benchmarked on the MNIST digit classification task, where it achieved $95.6\%$ accuracy.
In \cite{kappel2015network} it was shown that the variability of learning rules and weight parameters gives rise to a biologically plausible model of online learning. The intrinsic noise of synaptic weight changes drives a sampling process that can be used to exploit redundancies in the task solution space (see Fig.~\ref{fig:models}b for an illustration). This model was applied to unsupervised learning in spiking neural networks, and to closed-loop reinforcement learning problems \cite{kappel2018dynamic, kaiser2019embodied}. In \cite{yan2019efficient} this model was also ported to the SpiNNaker~2 neuromorphic many-core system.
\vspace{5mm}

\subsubsection{Biologically motivated mechanisms to combat forgetting in always-on learning scenarios}
Neuromorphic systems often operate in an environment where they are permanently on and learning a continuous stream of data. This mode of operation is quite different from most other machine learning applications that work with hand-labeled batches of training data. Always-on learning on a system with limited resources inevitably leads to situations where the system reaches the limits of its memory capacity and thus starts forgetting previously learned sensory experiences. Inspiration to overcome forgetting relevant information comes from biology. The mammalian brain seems to combat forgetting by actively protecting previously acquired knowledge in neocortical circuits \cite{cichon2015branch,pan2009survey,hayashi2015labelling,YangETAL:09,yang2014sleep}. When a new skill is acquired, a subset of synapses is strengthened, stabilized and persists despite the subsequent learning of other tasks \cite{YangETAL:09}.

A theoretical treatment of the forgetting problem was conducted in the \emph{cascade model} of Stefano Fusi and others \cite{fusi2005cascade,benna2016computational}. They could show that learning an increasing number of patterns in a single neural network leads unavoidably to a state which they called catastrophic forgetting. Trying to train more patterns into the network will interfere with all previously learned ones, effectively wiping out the information stored in the network. The proposed cascade model to overcome this problem uses multiple parameters per synapse that are linked through a cascade of local interactions. This cascade of parameters selectively slows down weight changes, thus stabilizes synapses when required and effectively combats effects of forgetting. A related model, that uses multiple parameters per synapse to combat forgetting was used in \cite{KirkpatrickETAL:17} (see also \cite{huszar2018note} for a recently introduced variation of the model). They used a Bayesian approach that infers a prior distribution over parameter values at each synapse. Synapses that stabilize during learning (converge to a fixed solution) will be considered relevant in subsequent learning and Bayesian priors help to maintain their values (see Fig.~\ref{fig:models}c for an illustration).
\vspace{5mm}

\subsubsection{Biologically motivated mechanisms to enhancing transfer and sensor fusion}
Distributed computing architectures at the edge need to make decisions by integrate information from different sensors and sensor modalities and they should be able best make use of the sensory information across a wide range of tasks. It is clearly not very efficient to learn from scratch when confronted with a new task. Therefore, to boost the performance of edge computing, we will consider two aspects of transferring information to new situations: transfer of knowledge between sensors (\emph{sensor fusion}), which has been treated in Section~\ref{sec:sensor-fusion}, and transfer of knowledge between multiple different tasks (\emph{transfer learning}).

\emph{Transfer learning} denotes the improvement of learning in a new task through the use of knowledge from a related task that has already been learned previously \cite{caruana1997multitask,torrey2010transfer}. This contrasts most other of today's machine learning applications that focus on one very specific task. In transfer learning, when a new task is learned, knowledge from previous skills can be reused without interfering with them. E.g. the ability to perform a tennis swing can be transferred to playing ping pong, while maintaining the ability to do both sports. The literature on transfer learning is extensive and many different strategies have been developed depending on the relationship between the different task domains (see \cite{weiss2016survey} and \cite{lu2015transfer} for systematic reviews). In machine learning a number of approaches have been applied to a wide range of problems, including classification of images \cite{long2017deep,duan2012learning,kulis2011you,zhu2011heterogeneous}, text \cite{wang2011heterogeneous, zhou2014heterogeneous, prettenhofer2010cross, zhou2014hybrid} or human activity \cite{harel2010learning}.

A very general approach to learn across multiple domains is followed in the \emph{learning to learn} framework of \cite{schmidhuber1992learning,schmidhuber1993neural}. Their model features networks that are able to modify their own weights through the network activity. These network are therefore able to tinker with their own processing properties. This approach has been taken to its most extreme form where a network leans to implement an optimization algorithm by itself \cite{andrychowicz2016learning}. This model consists of an outer-loop learning network (\emph{the optimizer}) that controls the parameters of an inner-loop network (\emph{the optimizee}). The training algorithm of the inner-loop network works on single tasks that are presented sequentially, whereas the outer-loop learner operates across tasks and can acquire strategies to transfer knowledge. This learning-to-learn framework was recently applied to SNNs to obtain properties of LSTM networks and use them to solve complex sequence learning tasks \cite{bellec2018long}. In \cite{bohnstingl2019neuromorphic} the learning-to-learn framework was also applied to a neuromorphic hardware platform.



\section{Signal processing for wearable devices on neuromorphic chip}
\label{sec:cmos}
Neuromorphic engineering is a branch of electrical engineering dedicated to the design of analog/digital data processors that aims to emulate biological neurons and synapses. It typically consumes less energy than conventional computing systems and presents additional properties, such as massively parallel event-based computation, distributed local memory and adaptation~\cite{indiveri2015memory, chicca2014neuromorphic}. 
This increasing interest in neuromorphic engineering shows that hardware \acp{SNN} are considered a key future technology with high potential 
in key application, such as the Edge of Computing, and wearable devices.

Neuromorphic technologies have sparked interest from universities~\cite{Furber2014, schemmel2020accelerated,  qiao2015reconfigurable, Moradi_etal17, neckar2018braindrop} and companies such as IBM~\cite{Merolla2014} and Intel~\cite{Davies_etal18}. 
In this Section, we will provide an overview of the neuromorphic platforms, that to the best of our knowledge were deployed for biomedical signal processing, showing promising results to be exploited in wearable devices.

\subsection{Neuromorphic processors}
\label{ssec:cmos}

\textbf{\emph{TrueNorth.}} 
TrueNorth~\cite{Merolla2014} is IBM's fully digital neuromorphic chip with one million neurons arranged in a tiled array of 4096 neurosynaptic cores enabling \emph{massive parallel processing}. 
Each core contains~13kB of \emph{local SRAM memory} to keep neurons and synapse's states along with the axonal delays and information on the fan-out destination.
There are 256 Leaky-Integrator and Fire (LIF) neurons implemented by time-multiplexing and 256 million synapses are designed in the form of SRAM memory.
Each core can support up to 256 fan-in and fan-out, and this connectivity can be configured such that a neuron in any core can communicate its spikes any other neuron in any other core.
\\
Thanks to the \emph{event-driven}, the co-location of memory and processing units in each core, and the use of low-leakage silicon CMOS technology, TrueNorth can perform 46 billion synaptic operations per second (SOPS) per watt for real-time operation, with 26 pJ per synaptic event. Its power density of 20 mW/cm$^2$ is about three orders of magnitude smaller than that of typical CPUs.

\textbf{\emph{SpiNNaker.}} 
The SpiNNaker machine~\cite{Furber2014}, designed by the University of Manchester, is a custom-designed ASIC based on \emph{massively parallel architecture} that has been designed to efficiently simulate large spiking neural networks. It consists of ARM968 processing cores arranged in a 2D array where the precise details of the neurons and their dynamics can be programmed into.
Although the processing cores are synchronous microprocessors, the \emph{event-based} aspect of SpiNNaker is apparent in its message-handling paradigm. A message (event) gets delivered to a core generating a request for being processed. 
The communications infrastructure between these nodes is specially optimized to carry very large numbers of very small packets, optimal for spiking neurons. 
\\
A second generation of SpiNNaker was designed by Technical University of Dresden~\cite{mayr2019spinnaker}. Spinnaker2 continues the line of dedicated digital neuromorphic chips for brain simulation increasing the simulation capacity by a factor $>10$ while staying in the same power budget (i.e. 10x better power efficiency). The full-scale SpiNNaker2 consists of 10 Million ARM cores distributed across 70000 Chips in 10 server racks.
This system takes advantage of advanced 22nm FDSOI technology node with Adaptive Body Biasing enabling reliable and ultra-low power processing. It also features incorporating numerical accelerators for the most common operations.

\textbf{\emph{Loihi.}} 
Loihi~\cite{Davies_etal18} is Intel's neuromorphic chip with many core processing incorporating on-line learning designed in 14\,nm  FinFET technology. The chip supports about 130000 neurons and 130 million synapses distributed in 128 cores. Spikes are transported between the cores in the chip using packetized messages by an asynchronous network on chip. It includes three embedded x86 processors and provides a very flexible learning engine on which diverse online learning algorithms such as \ac{STDP}, different 3 factor and trace-based learning rules can be implemented. The chip also provides hierarchical connectivity, dendritic compartments, synaptic delays as different features that can enrich a spiking neural network. The synaptic weights are stored on local SRAM memory and the bit precision can vary between 1 to 9 bits. All logic in the chip is digital, functionally deterministic, and implemented in an asynchronous bundled data design style.

\textbf{\emph{DYNAP-SE.}} 
DYNAP-SE implements a multi-core neuromorphic processor with scalable architecture fabricated using a standard 0.18 $\mu m$ CMOS technology~\cite{Moradi_etal17}. It is a full-custom asynchronous mixed-signal processor, with a fully asynchronous inter-core and inter-chip hierarchical routing architecture. Each core comprises 256 adaptive exponential integrate-and-fire  (AEI\&F) neurons for a total of 1k neurons per chip. Each neuron has a Content Addressable Memory (CAM) block, containing 64 addresses representing the pre-synaptic neurons that the neuron is subscribed to. 
Rich synaptic dynamics are implemented on the chip by using \ac{DPI} circuits ~\cite{Bartolozzi_Indiveri2007}.
These circuits produce EPSCs and IPSCs (Excitatory/Inhibitory Post Synaptic Currents), with time constants that can range from a few $\mu s$ to hundreds of $ms$. 
The analog circuits are operated in the sub-threshold domain, thus minimizing the dynamic power consumption, and enabling implementations of neural and synaptic behaviors with biologically plausible temporal dynamics. 
The asynchronous CAMs on the synapses are used to store the tags of the source neuron addresses connected to them, while the SRAM cells are used to program the address of the destination core/chip that the neuron targets. 

\textbf{\emph{ODIN/MorphIC.}} 
ODIN (Online-learning DIgital spiking Neuromorphic) processor occupies an area of only 0.086mm$^2$ in 28nm FDSOI CMOS \cite{frenkel20180}. It consists of a single neurosynaptic core with 256 neurons and 256$^2$ synapses. Each neuron can be configured to phenomenologically reproduce the 20 Izhikevich behaviors of spiking neurons \cite{Izhikevich2004}. The synapses embed a 3-bit weight and a mapping table bit that allows enabling or disabling Spike-Dependent Synaptic Plasticity (SDSP) locally \cite{brader2007learning}, thus allowing for the exploration of both off-chip training and on-chip online learning setups.
\\
MorphIC is a quad-core digital neuromorphic processor with 2k LIF neurons and more than 2M synapses in 65nm CMOS \cite{frenkel201965}. MorphIC was designed for high-density large-scale integration of multi-chip setups. The four 512-neuron crossbar cores are connected with a hierarchical routing infrastructure that enables neuron fan-in and fan-out values of 1k and 2k, respectively. The synapses are binary and can be either programmed with offline-trained weights or trained online with a stochastic version of SDSP.

\subsection{Biomedical signal processing on Neuromorphic hardware}
\label{sec:neuproc}
Table~\ref{tab:neurochips} shows the summary of neuromorphic processors described previously and in which biomedical signal processing applications were used. These works show promising results for always-on embedded biomedical systems. 

The first chip presented in this table is DYNAP-SE, used to implement \acp{SNN} for the classification or detection of \ac{EMG}~\cite{donati2018processing, donati2019discrimination} and \ac{ECG}~\cite{Bauer_etal19, Corradi_etal19} and to implement a simple spiking perceptron as part of a design to detect \ac{HFO} in human intracranial \ac{EEG}~\cite{Sharifshazileh_etal19}.
In particular, in~\cite{donati2018processing,Bauer_etal19} a spiking \ac{RNN} is deployed for \ac{ECG}/\ac{EMG} signal separation to facilitate the classification with a linear read-out. \ac{SVM} and linear least square approximation is used in the read out layer for \cite{Bauer_etal19,Corradi_etal19} and overall accuracy of $91\%$ and $95\%$ for anomaly detection were reached respectively. In \cite{donati2018processing}, the state property of the spiking \ac{RNN} on \ac{EMG} was investigated for different hand gestures. 
In~\cite{donati2019discrimination} the performance of a feedforward \ac{SNN} and a hardware-friendly spiking learning algorithm for hand gesture recognition using superficial \ac{EMG} was investigated and compared to traditional machine learning approaches, such as \ac{SVM}. Results show that applying \ac{SVM} on the spiking output of the hidden layer achieved a classification rate of $84\%$, and the spiking learning method achieved $74\%$ with a power consumption of about $0.05~mW$. The consumption was compared to state-of-the-art embedded system showing that the proposed spiking network is two orders of magnitude more power efficient~\cite{benatti2015versatile, montagna2018pulp}.
   
Recently, the benchmark hand-gesture classification was processed and compared on two other digital neuromorphic platforms, i.e. Loihi and ODIN/MorphIC~\cite{frenkel20180, frenkel201965}. A spiking \ac{CNN} was implemented on Loihi and a spiking \ac{MLP} was implemented on ODIN/MorphIC~\cite{Ceolini_etal20}.
Because of the properties of neuromorphic chips, on Loihi a late fusion was implemented combining the output from the spiking \ac{CNN} for vision, and the spiking \ac{MLP} for \ac{EMG} signals; While on ODIN/MorphIC hardware, the two spiking \acp{MLP} were fused in the last layer.
Due to the neuromorphic chip properties the Loihi implemented a late fusion of a spiking \ac{CNN}, for vision and a spiking \ac{MLP} for \ac{EMG} signals. In the ODIN/MorphIC system two spiking \acp{MLP} were fused in the last layer. 
The comparison with the embedded GPU was performed in terms of accuracy, power consumption,  and latency showing that the neuromorphic chips are able to achieve the same accuracy with significantly smaller energy-delay product, 30x and 600x more efficient for Loihi and ODIN/MorphIC, respectively~\cite{Ceolini_etal20}.
   
\begin{table*}[t]
	\small
	\centering
    \caption{Summary of neuromorphic platforms and biomedical applications}
	\vspace{0.2cm}
	\label{tab:neurochips}
	\renewcommand{\arraystretch}{1.2}
  \begin{tabular}
       {>{\centering\arraybackslash} m{3.7cm} 
       *1{ |>{\centering\arraybackslash} m{2.5cm}}
       *1{ |>{\centering\arraybackslash} m{2.5cm}}
       *1{| >{\centering\arraybackslash} m{2cm}}
       *1{| >{\centering\arraybackslash} m{2cm}} 
       *1{| >{\centering\arraybackslash} m{2.5cm}} 
       }
\hline
\toprule
	\textbf{Neuromorphic Chip}	& \textbf{DYNAP-SE} & \textbf{SpiNNaker} & \textbf{Loihi} & \textbf{TrueNorth} & \textbf{ODIN} \\ 
	\hline
	\hline
	\textbf{CMOS Technology}	& 180nm & ARM968, 130 nm & 14nm FinFET & 28nm & 28 nm FDSOI   \\ 
	\hline
	\textbf{Implementation}	& Mixed-signal & Digital & Digital ASIC & Digital ASIC & Digital ASIC  \\ 
	\hline
	\textbf{Energy per SOP}	& 17 pJ @ 1.8V & Peak power 1W per chip & 23.6 pJ @ 0.75V & 26 pJ @ 0.775 & 12.7 pJ@0.55V \\
	\hline
	\textbf{Size}	& 38.5 $mm^2$ & 102 $mm^2$ & 60 $mm^2$ & 0.093 $mm^2$ (core) & 0.086 $mm^2$  \\ 
	\hline
	\textbf{On-chip learning} & No & Yes (configurable) & Yes (configurable) & No & Yes (SDSP)\\ 
	\hline	
	\textbf{Applications}	& \ac{EMG}, \ac{ECG}, \ac{HFO} &  \ac{EMG} and \ac{EEG} & \ac{EMG} & \ac{EEG} and \ac{LFP} & \ac{EMG} \\
	\hline
	\end{tabular}
\end{table*}

\subsection{Encoding}
\label{ssec:encoding}
In \acp{SNN} a single spike by itself does not carry any information. However, the number and the timing of spikes produced by a neuron are important. Just as their biological counterpart, silicon neurons in neuromorphic devices produce spike trains at a rate that is proportional to their input current. At the input side, synapse circuits integrate the spikes they receive to produce analog currents, with temporal dynamics and time constants that can be made equivalent to their biological counterparts. The sum of all the positive (excitatory) and negative (inhibitory) synaptic currents afferent to the neuron is then injected into the neuron.

To provide biomedical signals to the synapses of the \ac{SNN} input layer, it is necessary to first convert them into spikes. A common way to do this is to use a delta-modulator circuit~\cite{Corradi_etal15,Sharifshazileh_etal19} functionally equivalent to the one used in the Dynamic Vision Sensor (DVS)~\cite{Lichtsteiner_etal2008}. 
This circuit, in practice, is an ADC that produces two asynchronous digital pulse outputs (UP or  DOWN) for every biosignal channel in the input. The UP (DOWN) spikes are generated every time the difference between the current and previous value exceeds a pre-defined threshold. The sign of the difference corresponds to the UP or DOWN channel where the spike is produced. This approach was used to convert \ac{EMG} signals, used in mixed-signal neuromorphic chips~\cite{donati2018processing, donati2019discrimination} and in digital ones~\cite{Behrenbeck_etal19, Ceolini_etal20}, \ac{ECG} signals~\cite{Corradi_etal19, Bauer_etal19}, and \ac{EEG} and \ac{HFO} ones~\cite{Corradi_etal15, Sharifshazileh_etal19}.
\\
\subsection{Adaptation in neuromorphic processor}

Local adaptation is an important aspect in extreme edge computing, specially when it comes to wearable devices. The current methods for training networks for biomedical signals rely on large datasets collected from different patients. However, when it comes to biological data, there is no ``one size fits all''. Each patient and person has their own unique biological signature. Therefore, the field of Personalized Medicine (PM) has gained lots of attention in the past few years and the online on-edge adaptation feature of neuromorphic chips can be a game changer for PM. 

As was discussed in Section \ref{sec:bio-algorithms}, there are lots of effort in designing spike-based online learning algorithms which can be implemented on neuromorphic chips.

Example of today's state of the art for on-chip learning are Intel's Loihi \cite{Davies_etal18}, DynapSEL and ROLLS chip from UZH/ETHZ~\cite{qiao2016scaling, qiao2015reconfigurable}, BrainScales from Heidelberg \cite{Schemmel2010} and ODIN from UC Louvain \cite{frenkel20180}. 
Intel's Loihi includes a learning engine which can implement different learning rules such as simple pairwise STDP, triplet STDP, reinforcement learning with synaptic tag assignments or any 3 factor learning rule implementation. DynapSEL, ROLLS and ODIN encompass the SDSP, also known as the Fusi learning rule, which is a form of semi-supervised learning rule that can support both unsupervised clustering applications and supervised learning with labels for shallow networks \cite{brader2007learning}. BrainscaleS chip implements the STDP rule. 
Moreover, Spinnaker 1 and 2 \cite{Furber2013,mayr2019spinnaker} can implement a wide variety of on-chip learning algorithms since their designs make use of ARM microcontrollers providing lots of configurability for the users.
\\
\subsection{Open challenges}
Generally, implementing on-chip online learning is challenging because of these two core reasons: 
locality of the weight update and weight storage. 

\textbf{\emph{Locality}} The learning information for updating the weights of any on-chip network should be locally available to the synapse since otherwise this information should be ``routed'' to the synapse by wires which will take a significant amount of area on chip.
The simplest form of learning which satisfies this requirement is Hebbian learning which has been implemented on a variety of neuromorphic chips forms of unsupervised/semi-supervised learning \cite{frenkel20180,qiao2015reconfigurable,Schemmel2010,qiao2016scaling}.  
However, Hebbian-based algorithms are limited in the tasks they can learn and to the best of our knowledge no large scale task has been demonstrated using this rule. 
Since gradient descent-based algorithms such as \ac{Backprop} has had lots of success in deep learning, there are more and more spike-based error \ac{Backprop} rules that are being developed as was discussed in Section \ref{sec:bio-algorithms}.  
These types of learning algorithms have recently been custom designed in the form of spike-based delta rule as back-bone of the \ac{Backprop} algorithm. For example, single layer implementation of the delta rule has been designed in  \cite{payvand_indiveri_19} and employed for \ac{EMG} classification \cite{donati2019discrimination}. Expanding this to multi-layer networks involves non-local weight updates which limits its on-chip implementation. Making the \ac{Backprop} algorithm local is a topic of on-going research which we have discussed in Section \ref{sec:bio-algorithms}. Recently, a multi-layer perceptron error-triggered learning architecture has been proposed to overcome the non-locality of multi-layer networks solving the spatial credit assignment problem on chip \cite{payvand2019error,Payvand_etal_2020_errormlp}
    
\textbf{\emph{Weight storage}} The ideal weight storage for online on-chip learning should have the following properties: 
(i) non-volatility to keep the state of the learnt weights even when the power shuts down to reduce the time and energy footprints of reloading the weights to the chip. (ii) Linear update which allows the state of the memory to change linearly with the calculated update. (iii) Analog states which allows a full-precision for the weights. 
Non-volatile memristive devices have been proposed as a great potential for the weight storage and there is a large body of work combining the CMOS technology with that of the memristive devices to get the best of two worlds.

In the next Section we provide a thorough review on the state of the art for the emerging memory devices and the efforts to integrate and use them in conjunction with neuromorphic chips.

\section{Memristive devices and computing}
\label{sec:memristive}
The severe power and area constraints under which a neuromorphic processor for edge computing must work opened ways towards the investigation of beyond-CMOS solutions. Despite still at the dawn of its technological development, memristive devices have been drawing attention in the last decade thanks to their scalability, low-power operation, compatibility with CMOS chip power supply and CMOS fabrication process, and volatile/non-volatile properties. In Section~\ref{subsec:memdevices}, we will introduce memristive devices and the properties that are appealing for adaptive extreme edge computing paradigms. In Section~\ref{subsec:memcomputing}, we will explore the role of memristive devices in neuromemristive systems and give examples of possible applications. In Section~\ref{subsec:memchallenge}, we will discuss the current challenges and the future perspectives of memristive technology.

\subsection{Conventional and wearable memristive devices}
\label{subsec:memdevices}

\begin{figure}
    \centering
    \includegraphics[width=0.9\textwidth]{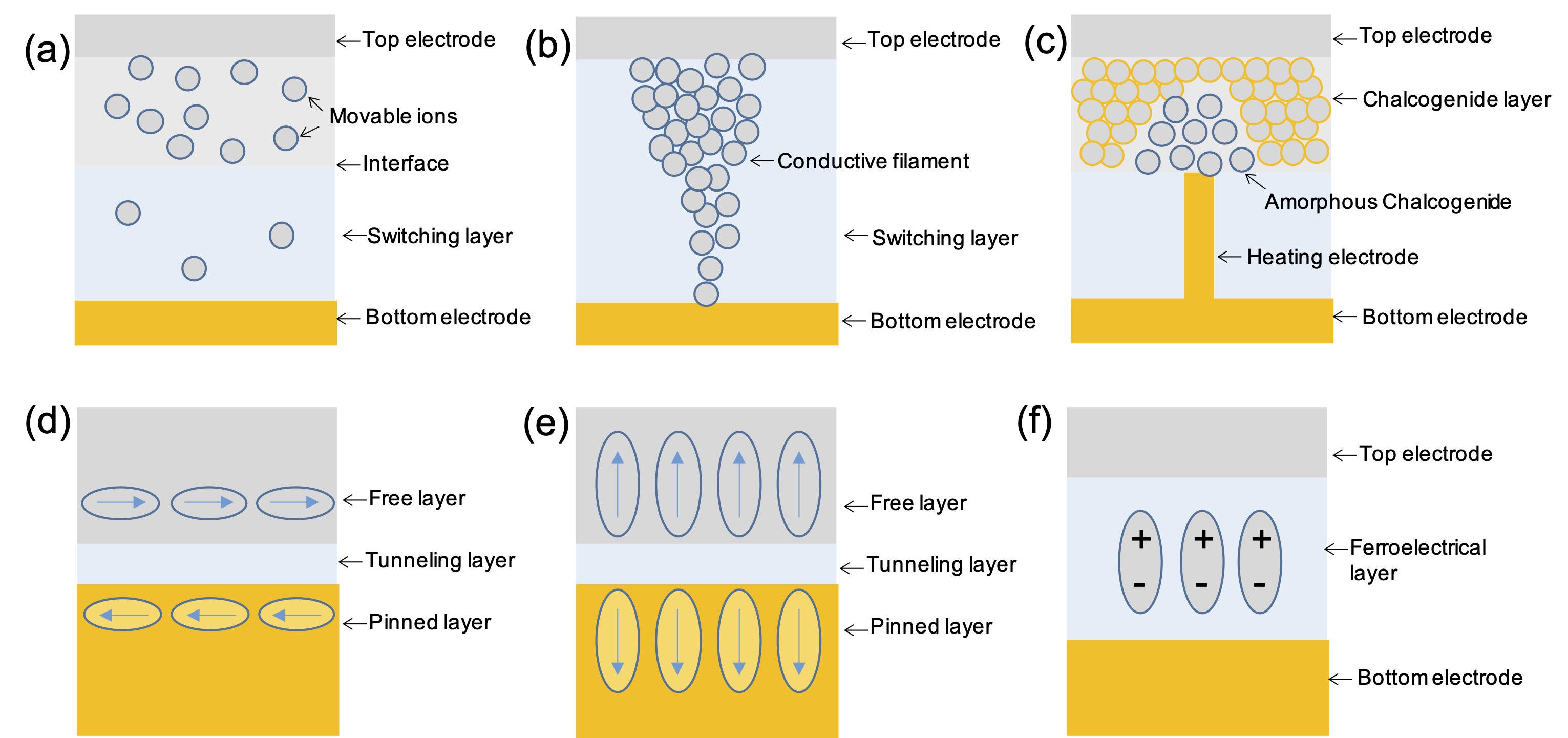}
    \caption{Memristive devices for neuromorphic computing. (a) Interface type RRAM device; (b) Filamentary RRAM device; (c) Phase change memory device; (d) MRAM device with in-plane spin polarization; (e) MRAM device with perpendicular spin polarization; (f) FTJ device. }
    \label{fig:mem_device}
\end{figure}

Memristive devices, as the name suggested, are devices which can change and memorize their resistance states. They are usually two-terminal devices, however, can be implemented with various physical mechanisms, resulting in versatile existing forms, e.g. resistive random access memory (RRAM, Fig.~\ref{fig:mem_device}a and ~\ref{fig:mem_device}b) (\cite{Ielmini2018}), phase change memory (PCM, Fig.~\ref{fig:mem_device}c) (\cite{Zhang2019}), magnetic random access memory (MRAM, Fig.~\ref{fig:mem_device}d and Fig.~\ref{fig:mem_device}e) (\cite{Miron2011}), ferroelectric tunneling junction (FTJ, Fig.~\ref{fig:mem_device}f) (\cite{Wen2013}), etc. The resistance memory of these devices can mimic the memory effect of the basic components of biological neural system, while the resistance changing can mimic the plasticity of biological synapse. Facilitated with their simplicity of two-terminal configuration and scalability to nanoscale, they are inherently suitable for the hardware implementation of brain-inspired computation materializing an artificial neural network, i.e. neuromorphic computation (\cite{Jo2010,Wang2016c}). 

This notation, in recent years, has incited wide investigations on the various memristive devices and on their applications in neural network learning and recognition, or, in short, memristive learning (\cite{Ohno2011,Kuzum2012,Yang2013,Alibart2013,Eryilmaz2014,Ambrogio2018}). The memristive learning can enable energy efficient and low latency information process within a reduced size of systems abandoning the conventional von-Neumann architecture. Among other benefits, this will also make it possible to process information where they are acquired, i.e. within sensors, and reduce the bandwidth needed for transferring the sensor data to data center, accelerating the coming of the era of Internet-of-Things (IOT). Table \ref{tab:memdev} summarizes the key features of the main memristive device technologies for neuromorphic / wearable applications in terms of cell area, electrical characteristics, main advantages and challenges. It is worth noticing that some figures of merit in this context are radically different with respect to standard memory requirements. Indeed, while in the memory scenario higher read currents enable faster reading speed, in neuromorphic applications currents as low as possible are preferred, since the current is a limiting factor for neurons' fan-out. Similarly, SET and RESET times should be as fast as possible in memory applications, while in our applications this requirement can be relaxed thanks to the lower operating frequency of the neurons (20\,Hz to 100\,Hz). Moreover, the number achievable conductance levels has to be increased (\cite{ielmini2020AdvIntellSys}). Some non-idealities which are usually detrimental for memory applications, for instance stochasticity of switching parameters, are even beneficial for the neural networks.

\begin{table*}[t]
    \small
	\centering
    \caption{Key features of non-volatile memristive devices.} 
    \label{tab:memdev}
	\vspace{0.2cm}
	\renewcommand{\arraystretch}{1.2}
    \begin{tabular}{|>{\centering\arraybackslash} m{2.7cm} 
       *1{ |>{\centering\arraybackslash} m{3.2cm}}
       *1{ |>{\centering\arraybackslash} m{3.2cm}}
       *1{| >{\centering\arraybackslash} m{3.2cm}}
       *1{| >{\centering\arraybackslash} m{3.2cm}|} 
       }
       \toprule
         \hline
         & \textbf{RRAM} & \textbf{PCM} & \textbf{MRAM} & \textbf{FTJ} \\
         \hline
         \hline
         \textit{Cell area [min. feature size]} & $4F^2$ \cite{irds2020} & $4F^2$ \cite{irds2020} & $9F^2$ (\cite{rho2017ISSCC}) & $4F^2$ \cite{irds2020} \\
         \hline
         \textit{Retention} & $>$10 years (\cite{goux2014VLSI}) & $>$10 years (\cite{cheng2012IEDM}) & $>$10 years (\cite{golonzka2018IEDM}) & $>$10 years (\cite{udayakumar2013IMW}) \\
         \hline
         \textit{Endurance} & $10^{12}$ (\cite{kim2011VLSI,lee2011NatMat}) & $10^{11}$ (\cite{kim2010VLSI}) & $10^{12}$ (\cite{saida2017TED}) & $>10^{15}$ (\cite{udayakumar2013IMW}) \\
         \hline
         \multirow{2}{*}{\textit{SET / RESET time}} & 100\,ps (\cite{torrezan2011Nanotech}) & $>$100\,ns, 10\,ns & 20\,ns (\cite{jan2018VLSI}) & 30\,ns, 30\,ns \\
            & 85\,ps (\cite{choi2016AdvFunctMat}) & (\cite{irds2020}) & 3\,ns (\cite{kitagawa2012IEDM}) & (\cite{francois2019IEDM}) \\
         \hline
         \textit{Read current} & 100\,pA (\cite{luo2016Nanoscale}) &  25\,$\mu$A (\cite{desandre2010ISSCC}) & 20\,$\mu$A (\cite{kitagawa2012IEDM}) & 0.8\,nA (\cite{bruno2016AdvElectrMat}, device diameter 300\,nm)  \\ 
         \hline
         \textit{Write energy per bit} & 20\,fJ (\cite{kang2015NanoEnergy}) & $\sim$100\,fJ (\cite{xiong2011Science}) & 90\,fJ (\cite{kitagawa2012IEDM}) & $<$10\,fJ (\cite{francois2019IEDM}) \\
         \hline
         \textit{Main features} & Scalability, speed, low energy & Scalability, multilevel, low voltage & Endurance, low power & Endurance, low power, speed \\
         \hline
         \textit{Challenges} & Variability & RESET current, temperature stability, resistance drift & Density, scalability, variability & Scalability \\
         \hline
    \end{tabular}
\end{table*}

In addition to the commonly referred non-volatile type of memristive switching, the RRAM device can also show volatile behavior, which usually occurs when active materials such as silver or copper are used as electrode. 
The relatively long retention time of the volatile behavior (tens of milliseconds to seconds) is then found to be similar to the timescale of short term memory, and naturally was proposed to mimic the short term memory effect of biological synapses (\cite{ZhongruiWang2017,WeiWang2019ted2,covi2019ICECS}). 

Although most researches on memristive devices are carried on rigid silicon substrates, the simple 
structure of memristive devices can also be realized on flexible substrates (\cite{Shi2020}), which opens new interesting possibilities for realizing local computation within wearable devices (\cite{Shang2017,Dang2019}). 

\subsection{Memristive devices for neuromorphic computing}
\label{subsec:memcomputing}

\subsubsection{Memristive neural components}
\label{subsec:memcomponents}
\begin{figure}
    \centering
    \includegraphics[width=0.9\textwidth]{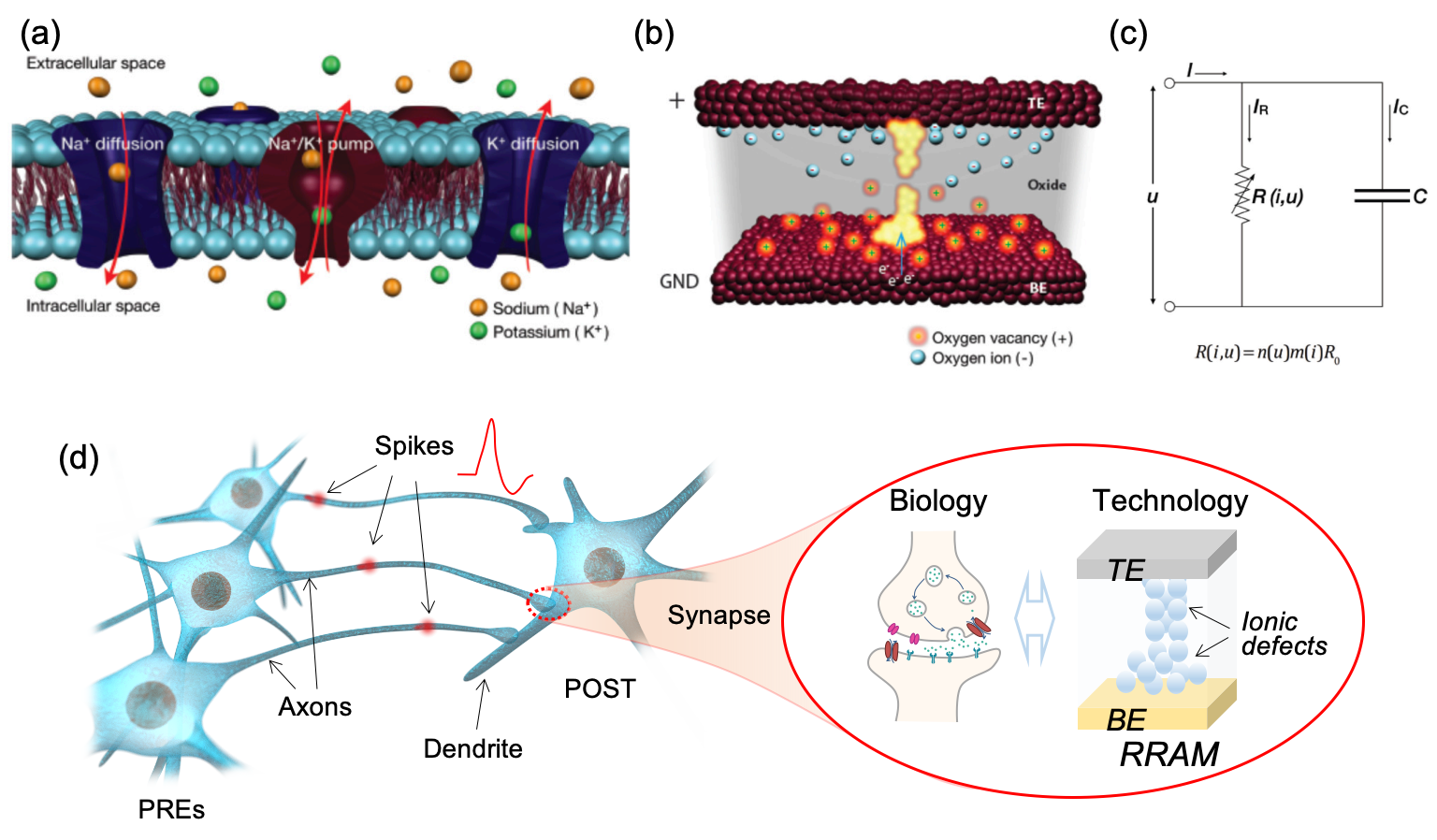}
    \caption{Memristive devices as synapse or neuron for neuromorphic computing. (a)-(c) memristive device act as threshold device for the firing function of biological neuron (\cite{Mehonic2016front}, reproduced under the CC BY license). (d) Conceptual illustration of memristive device as artificial synapse for brain-like neuromorphic computing (\cite{WeiWang2018sa}, reproduced under the CC BY-NC license). }
    \label{fig:mem_comp}
\end{figure}

As mentioned in Section~\ref{subsec:memdevices}, the primary function of memristive devices is the usage as synaptic devices to implement the memory and plasticity of biological synapses. However, there are increasing interests for these devices to be utilized to implement nanoscale and artificial neurons.

On the neuron side, the memristive device gradual internal state change and its consequently abrupt switching closely mimic the integrate-and-fire behavior of biological neurons (\cite{Mehonic2016front,Tuma2016,suresh2019ICECS}, Fig.~\ref{fig:mem_comp}a-c). Due to the sample structure and nanometer level scalability, memristive neurons can be much more compact than current CMOS neurons which might consist of current sensor, analog-to-digital converter (ADC), and analog-to-digital converter (DAC), and capacitors, all of which are expensive to implement in current CMOS technology in terms of area and/or power consumption (\cite{kwon2018JAP}). The implementation of memristive neurons will also enable full memristive neuromorphic computing (\cite{wang2018NatElectr}), which promises further increases in the integration of the hardware neuromorphic computing.

On the synaptic side, the key feature of the biological synapses is their plasticity, i.e. tunable weight, which can be generally implemented by resistance or conductance modification in the memristive devices (Fig.~\ref{fig:mem_comp}d). 
Fundamental learning rules based on \ac{STDP} have already been widely explored (\cite{Kuzum2012,WangZhongqiang2015,covi2016FrontNeurosci,mulaosmanovic2017VLSI,covi2018JPhysD}). Spatial spiking pattern recognition (\cite{Pedretti2017}), spiking co-incidence detection (\cite{Prezioso2018,Sebastian2017}), and spatial-temporal correlation (\cite{WeiWang2018sa,WeiWang2019fd}) has been reported recently. Synaptic metaplasticity, such as paired-pulse facilitation, can also be achieved via various device operation mechanism (\cite{Zhu2017,ZhongruiWang2017,Wu2018}).
\\

\subsubsection{Memristive neural network architectures}

There are generally two approaches for a hardware neuromorphic system implementing memristive devices as synapses: (i) deep learning accelerator, accelerating the artificial neural network computing with multiple layer and error back-propagation, as well as it's variations, like convolutional neural network, recurrent neural network, etc.; (ii) brain-like computing, attempting to closely mimicking the behaviors of biological neural system, like spike representation (Fig.~\ref{fig:mem_comp}d) and collective decision making behavior. In the deep learning accelerator approach, on-line training places more requirements for the memristive synapses. For instance, linear and symmetrical weight update is crucial for the on-line training (\cite{Burr2015,Ambrogio2018}), while off-line training ignores it since the synaptic weight can be programmed to the memristive device with fine tuning and iterative verify (\cite{Yao2020}). 

Collective decision making is an important feature of the brain computing, which requires high parallelism and, consequently, low current devices. For instance, this feature is the essential for Hopfield neural network (\cite{Hopfield1982}), cellular neural network (\cite{Duan2015}), and coupled oscillators (\cite{romera2018Nature}). In the Hopfield neural network, the system automatically evolves to its energy minimization points leading the functionality of associative memory. The use of Hopfield like recurrent neural networks (RNNs) with memristive devices has already been successfully demonstrated in a variety of tasks (\cite{milo2017IEDM,WangYanghao2020}). As an example of memristive based coupled oscillator network, \cite{ignatov2017SciAdv} used a network of self-sustained van der Pol oscillators coupled with oxide-based memristive devices to investigate the temporal binding problem, which is a well known issue in the field of cognitive neuroscience. In this experiment, the network is able to emulate an optical illusion which shows two patterns depending on the influence of attention. This means that the network is able to select relevant information from a pool of inputs, as in the case of a system collecting signals from multiple sensors.
\vspace{6mm}
\subsubsection{Applications of memristive neural networks}
At present, memristive technology has been mainly used in relatively simple networks with Hebbian-based learning algorithms. 
However, more recently, systems able of solving different tasks, such as speech recognition (\cite{park2015SciRep}), and exploring different architectures and learning algorithms are being investigated. In particular, the benefits of exploiting sparsity, mentioned in Section~\ref{sec:pruning}, are demonstrated for feature extraction and image classification in networks trained with stochastic gradient descend and winner-take-all learning algorithms (\cite{sheridan2016TNNLS}), as well as in hierarchical temporal memory, which does not need training (\cite{krestinskaya2018AnICSigProc}).

In the latest years, memristive devices have been used in applications closer to biology, enabling hybrid biological-artificial systems (\cite{serb2020SciRep}) and investigating biomedical applications, ranging from speech and emotion recognition (\cite{saleh2015CISDA}) to biosignal (\cite{kudithipudi2016FrontNeurosci}) and medical image (\cite{zhu2017Neurocomp}) processing. 
Finally, an interesting application is the one of memristive biosensors, which \cite{tzouvadaki2018ISCAS} used to implement a system for cancer diagnostic. The innovative use of memristive properties was demonstrated in hardware and opens the way to a broader use of memristive technology where sensors and computing co-exist in the same system or, possibly, in the same device.








\subsection{Open challenges and future work}
\label{subsec:memchallenge}

\subsubsection{Device non-idealities}

Implementation of mainstream deep learning algorithms with \ac{Backprop} learning rule and memristive synapses imposes some requirements for the memristive device, including linear current-voltage relation for reading, analog conductance tuning, linear and symmetric weight update, long retention time, high endurance, etc. (\cite{Gokmen2016}). However, no single device can fulfill all these requirements simultaneously. 

Various techniques have been proposed to compensate the device non-idealities. For instance, to compensate the non-linear current-voltage relation for reading, fixed read voltage with variable pulse width or pulse number can be used for synaptic weight reading, and the readout is represented by the charge accumulation in the output nodes (\cite{Cai2019}). Linear and symmetric weight update is crucial for accurate online learning of a memristive multilayer neural network with \ac{Backprop} learning rule (\cite{Burr2015}). However, PCM devices usually only show gradual switching in set direction (weight potentiation), while RRAM devices show gradual switching in reset direction (weight depression). To achieve linear and symmetric weight update, differential pair with two of these devices are usually used. For a differential pair with two PCM devices, the potentiation is achieved by applying set pulses on the positive part and the depression is achieved by applying set pulses on the negative part, thus gradual weight update in both potentiation and depression can be achieved. To further enhance the linearity of weight update, a minor conductance pair consisting of capacitors can be used for frequent but smaller weight update, and finally transferred to the major pair periodically (\cite{Ambrogio2018}). Another option to improve device linearity is limiting the device dynamic range in a region far from saturation and where the weight update is linear \cite{woo2016EDL,wang2016Nanoscale}.

In addition to mitigate the non-idealities of memristive devices, more and more research efforts are made to exploit these non-idealities for brain-like computations. For instance, the stochasticity or noise in reading of memristive device can be used for the probability computation for restricted Boltzmann machine (\cite{Mahmoodi2019}), or escape for local minimization points in a Hopfield neural network (\cite{Cai2020}). The Ag filament based resistive switching device shows short retention time and high switching dynamics, thus was proposed for reservoir computing (\cite{Midya2019a}) and spatiotemporal computing (\cite{WeiWang2019ted2}) to process time-encoded information. 

\subsubsection{Co-integration of hybrid CMOS-memristive neuromorphic systems}
The main steps to be taken to exploit the full potential of an \ac{ASIC} for end-to-end processing system go through the integration of memristive devices and sensors with CMOS technology. Indeed, the works presented so far are based either on simulations or on real device data, or on memristive chips interfaced with some standard digital hardware. Despite integration of CMOS technology has been demonstrated for non-volatile resistive switching devices already at a commercial level (\cite{scharlotta2014IIRW,hayakawa2015VLSI}), the design of co-integrated memristive-based neuromorphic processors is still under development. We envisage a three-phase process to achieve a fully integrated system.

The first one is the co-integration of non-volatile memristive devices with some peripheral circuits (\cite{hirtzlin2020FrontNeurosci}) and to implement some logic and multiply-and-accumulate (MAC) operations (\cite{chen2019NatElectr}), which reaches the maturity with the demonstration of a fully cointegrated \ac{SNN} with analog neurons and memristive synapses (\cite{valentian2019IEDM}). The second phase is the co-integration of different technologies. Despite this approach results in higher fabrication costs, it presents several advantages in terms of system performance, which can be more compact and potentially more power efficient. In particular, the co-integration of non-volatile and volatile memristive devices can lead to a fully memristive approach. As an example, \cite{wang2018NatElectr} exploit volatile memristive devices to emulate stochastic neurons and non-volatile memristive devices to store the synaptic weights on the same chip, thus demonstrating the feasibility and the advantages of the dual technology co-integration process. Eventually, the final step which has to be taken in the development of a dedicated \ac{ASIC} for wearable edge computing is the co-integration of sensors and memristive-based systems. \cite{shulaker2017Nature} tackled this challenge by designing and fabricating a gas sensing system able of gas classification. The system uses RRAM arrays as memory, Carbon Nanotube field effect transistor (CNFET) for computation and gas sensing, both 3D monolithically integrated on CMOS circuits, which carry out computation and allow memory access. 



\subsubsection{Learning with memristive devices}

Adaptability is a feature of paramount importance in smart wearable devices, which need to be able to learn the unique feature of their user. This calls for the implementation of lifelong learning paradigms, i.e. the ability of continuously learning new features from experience. Typically, a network has a limited memory capacity dependent on the network size and architecture. Once the maximum number of experiences is recorded, new features learned will erase old ones, thus originating the phenomenon of catastrophic forgetting.

The problem of an efficient implementation of continual learning has been thoroughly investigated (\cite{parisi2018arxiv}). In the current scenario, a dichotomy exist between backprop-based \ac{ANN}s, which have very high accuracy but a limited memory capacity, and brain-inspired \ac{SNN}s, which feature higher memory capacity thanks to their higher flexibility, but at the cost of lower accuracy. Models used to overcome forgetting are described in Section~\ref{sec:model-challenge}.
The use of memristive devices in such networks is still an open point. It is possible that memristive device will be beneficial to increase the network capacity (\cite{brivio2018Nanotech}) at no extra computational cost thanks to their slow approach to the boundaries (\cite{Frascaroli2018}), but so far this topic is still quite unexplored. An interesting approach is proposed by \cite{munozmartin2019JESSDCC}, where the key strengths of supervised convolutional \ac{ANN}s, unsupervised \ac{SNN}s, and memristive devices are combined in a single system. The results indicate that this approach is robust against catastrophic forgetting, whilst reaching 93\% accuracy when tested with both trained and non-trained classes.


\section{Discussion and Conclusions}
\label{sec:discussion}

In this study, we presented the state-of-the-art core elements that enable the development of wearable devices with extreme edge adaptive computing capability. Various sensors that can collect different bio-signals from the human body are investigated. There is a variety of sensing specifications in terms of size, resolution, mechanical flexibility and output signals need to be considered along with their analogue readout circuit at a limited amount of power consumption. However, when the real-time processing of these signals is deployed on edge, severe constraints raise in terms of power efficiency, fast response times, and accuracy in the data classification. The widely-used solution is to find a trade-off between the energy and computational capacity, or send the data to the cloud. However, these strategies are not ideal and slow down the development of wearable smart sensing. To meet all the requirements, the development of a platform needs to be optimized in synergy with the other elements and every aspect of the design, from the learning algorithms to the architecture.

In particular, continual learning is required for adaptive wearable devices. In this respect, brain-inspired algorithms promise to be valid alternatives to standard machine learning approaches such as Backprop and BPTT. The exploitation of sparsity in network connectivity increases the power efficiency by optimizing the use of the available memory. However, the problem of algorithmic robustness to non-ideal hardware (such as noise and variability) and the problems of forgetting and information transfer between tasks still persist and have to be solved in combination with neuromorphic and emerging technologies. \ac{SNN}s are conceptually ideal for low-power in-memory computing. Their event-based approach together with the use of analog subthreshold circuits to reproduce biological timescales, allows fast response times of the network while enabling smooth real-time processing of data. The encoding of the incoming signals into spikes is however still challenging. Moreover, a fully \ac{CMOS}-based approach has two major technological issues. First, the synaptic weight is usually stored in \ac{SRAM}s, which hold the state only in the presence of a power supply. Second, capacitors used to implement biological time constants are massive and may consume up to 60\% of the chip area. Memristive technology can be beneficial in this respect. Non-volatile devices can potentially replace \ac{SRAM}s and volatile devices offer a compact alternative to \ac{CMOS} capacitors. Besides low-power operation in a small footprint, memristive devices also offer noisy properties, which -- if exploited in the right way -- might facilitate the implementation of stochastic learning algorithms. However, the technology is still at its infancy and fabrication processes are still under development, yielding high device variability, which makes it difficult to produce reliable multi-bit memory.

In summary, the ultimate goal towards smart wearable sensing with edge computing capabilities relies on a bespoke platform consists of embedding sensors, front-end circuit interface, neuromorphic processor and memristive devices. This platform requires high-compatibility of existing sensing technologies with CMOS circuitry and memristive devices to move the intelligent algorithm into the wearable edge without significantly increase the cost in energy. New solutions are needed to enhance the performance of local adaptive learning rules to be competitive with the accuracy of Backprop. Novel encoding techniques to allow streamless communication from sensors to neuromorphic chip have to be developed and flanked by efficient event-based algorithms. So far there is not a uniquely ideal solution, but we envisage that a holistic approach where all the elements of the system are co-designed as a whole is the key to build low-power end-to-end real-time adaptive systems for next-generation smart wearable devices.


\section*{Conflict of Interest Statement}

The authors declare that the research was conducted in the absence of any commercial or financial relationships that could be construed as a potential conflict of interest.

\section*{Author Contributions}
All the Authors equally contributed to the manuscript, actively participating to the discussions and to the writing. The main contributors for each Section are as follows: X.L. and H.H. -- wearable sensors; D.K. -- biologically plausible models; M.P. and E.D. -- signal processing and neuromorphic computing. E.C. and W.W. -- memristive devices. E.C. led and coordinated the cooperative writing and all discussions.

\section*{Funding}
This work was partially supported by the UK EPSRC under grant EP/R511705/1. E.C. and M.P. acknowledge funding by the European Union‘s Horizon 2020 research and innovation programme under grant agreement No 871737.

\section*{Acknowledgments}
The Authors would like to thank Prof. Thomas Mikolajick and Dr. Stefan Slesazeck for useful discussion on ferroelectric and memristive devices.



\bibliography{refs.bib,test.bib}

\end{document}